\documentclass{easychair}

\usepackage{supertabular}
\usepackage{graphicx}
\usepackage{longtable}
\usepackage{amsmath}
\usepackage{tabularx}

\usepackage{caption}
\usepackage{subcaption}

\DeclareUnicodeCharacter{2212}{-}   

\usepackage{chronology}
\usepackage{algorithmic}
\usepackage[]{algorithm2e}
\usepackage{amssymb}
\usepackage{float}
\usepackage{threeparttable}

\usepackage{array}
\usepackage{subcaption}
\usepackage{multirow}
\usepackage{rotating}
\usepackage{pdflscape}

\usepackage{amssymb}

\begin{document}

%
\title{Adaptive Probabilistic Model for Energy-Efficient Distance-based Clustering in WSNs (Adapt-$P$): \\A LEACH-based Analytical Study}

%
\titlerunning{Adaptive Probabilistic Model for Energy-Efficient Clustering}

\volumeinfo
	{\emph{Journal of Wireless Mobile Networks, Ubiquitous Computing, and Dependable Applications}}
	{5}                         
	{4 (December)}                         
	{1}                         




\author{\\
Husam Suleiman${}^{}$\thanks{Corresponding author: Department of Electrical Engineering, Applied Science Private University, Amman, Jordan, Tel: +962-6-560-9999
}~~and Mohammad Hamdan${}^{}$\\
${}^{}$Applied Science Private University, Amman, Jordan\\
\url{h_suleiman@asu.edu.jo},~~~\url{mo_hamdan@asu.edu.jo}
}


%

%

\authorrunning{Suleiman and Hamdan}

\maketitle

%
\begin{abstract}
\noindent
Network lifetime and energy consumption of data transmission have been primary Quality of Service (QoS) obligations in Wireless Sensor Networks (WSNs). The environment of a WSN is often organized into clusters to mitigate the management complexity of such obligations. However, the distance between Sensor Nodes (SNs) and the number of clusters per round are vital factors that affect QoS performance of a WSN. The designer's conundrum resolves around the desire to sustain a balance between the limited residual energy of SNs and the demand for prolonged network lifetime. Any imbalance in controlling conflicting objectives may result in either QoS penalties due to draining SN energies which leaves WSN's environment unevenly covered, or an over-cost environment that is significantly difficult to distribute and operate.
Low-Energy Adaptive Clustering Hierarchy (LEACH) is a well-known distributed clustering algorithm proposed to tackle such difficulties. Multiple LEACH-based algorithms have also been proposed to enhance QoS requirements. Proposed algorithms typically focus on residual energies of SNs to compute a probability function that selects cluster-heads. Some algorithms form an optimal energy-efficient path toward a destination SN. Nevertheless, these algorithms do not consider variations in network's state at run-time. Such a state changes in an adaptive manner according to existing network structures and conditions. Thus, cluster-heads per round are not elected adaptively depending on the state and distances between SNs. To tackle such complications, this paper proposes an energy-efficient adaptive distance-based clustering called Adapt-$P$, in which an adaptive probability function is developed to formulate clusters. A near-optimal distance between each cluster-head and its cluster-members is formulated so that energy consumption of the network is mitigated and accordingly the network lifetime is maximized. Distances between and residual energies of SNs are employed to obtain a maximum number of cluster-heads to be elected per round. The cluster-head selection probability is adapted at the end of each round based on the maximum number of cluster-heads permitted per round found a priori and the number of existing alive SNs in the network. The Adapt-$P$ based algorithms proposed in this paper improve the performance of LEACH algorithm in term adaptivity and network lifetime.
\end{abstract}


\section{Motivation}

The implementation of a WSN is influenced by several design factors such as scalability, fault tolerant, topological layout, transmission media, and power consumption~\cite{SALEH2021, Alyousuf2019, Thi2019}. Fault tolerant, for example, ensures that a failure of a SN caused by a lack of power or physical damage should not affect the rest of SNs or severely degrade the overall performance. Since a WSN could be in the order of hundreds or thousands of SNs, scalable mechanisms are to be employed to help WSNs adapt to number and density of SNs~\cite{sharma2019heterogeneity, Guo2020, Shilpi2020}. Because each SN has limited power resources and may not be rechargeable, effective power-saving mechanisms are needed to extend network lifetime~\cite{Kirsan2019, Chen2019, Zhen2019}. Therefore, energy efficiency and network lifetime are primarily among the most important QoS obligations to consider in the design of a WSN~\cite{EACA2021, Pandey2021}. There is a need to formalize energy-efficient strategies that preserve SNs' residual energy, reduce the overall energy consumption, and consequently prolong the network lifetime~\cite{chan2020hierarchical, manuel2020optimization, Ateeq2020}.

Energy-efficient algorithms are proposed to address topological complications and energy concerns in WSNs. Heinzelman et al.~\cite{Heinzelman} present LEACH protocol that specifically merges ``clustering and media access with application-specific data aggregation'' to improve network lifetime, latency, and application quality. LEACH includes a distributed clustering technique that evenly rotates the cluster-head between SNs to avoid power depletion of a specific SN over others. They analytically compute the number of clusters expected per round to minimize the total energy consumption. 

Disjoint clustering is accordingly an efficient mechanism effectively employed in WSNs. It is utilized to achieve scalability, stabilize network topology, minimize energy consumption, aggregate sensor data, and alleviate congestion rate. A cluster-head would potentially limit the scope of communication between cluster-members and Base Station (BS), as well as coordinate the channel time among its cluster-members so that the whole network is not burden with extra redundant information. Periodic clustering distributes energy evenly among all SNs, which consequently increases the energy efficiency and extends the network lifetime. With respect to energy consumption, a cluster-member consumes energy when it transmits data to its corresponding cluster-head or directly to BS. On the other side, a cluster-head consumes energy when data is received from its cluster-members, and consumes more energy when data is aggregated and transmitted to the BS.

However, clustering approaches in the literature mostly focus on improving the process of selecting cluster-heads, either randomly or according to SN's residual energy. Some approaches form the best possible energy path toward BS such that network lifetime is increased. Nevertheless, such approaches do not account for variations in network state at run-time. In fact, the state varies often because of frequent changes in network configurations, and thus cluster-heads would not potentially be elected depending on existing conditions of the network and distances between SNs.

To tackle such difficulties, this paper presents a LEACH-based adaptive clustering algorithm in which an adaptive probability function is employed to formulate clusters. A near-optimal distance between each cluster-head and its corresponding cluster-members is formalized such that the overall energy consumption of the network is mitigated and accordingly the network lifetime is maximized. Distances between SNs and residual energy of each SN in the network are utilized to compute the maximum number of cluster-heads elected per round. The cluster-head selection probability is adapted at the end of each round according to the maximum number of cluster-heads permitted per round found in advance and the number of alive SNs currently interacting with the networking environment.

The proposed adaptive clustering scheme helps business organizations and building automation support real-time visibility and remote monitoring on their system operational data, so as to meet a high-level of energy-efficiency and cost-saving. The design is robust to topology changes and can be utilized to tackle interoperability between heterogeneous SNs, so that to it manages security and privacy challenges incurred from a large volume of enterprise data reported from the working environment.

\section{Contributions}
\label{sec:contr} 

The communication and computation energy model of the network is mathematically analyzed and examined, which is structured based on LEACH algorithm. An optimal distance ($d_{opt}$) between each cluster-head and its cluster-members is formulated, found by deriving the total energy consumption with respect to the distance. The $d_{opt}$ mitigates the total energy consumption of receiving, aggregating, and transmitting data, which as a result prolongs the network lifetime. Thus, the $d_{opt}$ is utilized to formulate the maximum number of clusters permitted per round, $\kappa_{\!{max}}$. As a result, the cluster-head selection probability is adapted at the end of each round based on $\kappa_{\!{max}}$ found a priori and number of alive SNs $\zeta$ currently active in the network. Additionally, distance to and residual energy of cluster-heads are both considered by each SN to select its potential cluster-head. Each SN joins a cluster-head that holds the maximum energy-distance ratio, $\nolinebreak{\digamma = \frac{E^{\alpha}_{res}}{d^{\beta}_{{C\!H}}}}$, where $E_{res}$ is the residual energy of cluster-head, $d_{{C\!H}}$ is the distance between SN and cluster-head, and $\alpha$ \& $\beta$ are constants. Contributions of this paper are summarized as follows:
\begin{itemize}
  \item Formulating $d_{opt}$ between each cluster-head and its cluster-members such that network lifetime is prolonged and the overall energy consumption is minimized.
  \item Formulating $\kappa_{\!{max}}$ based on $d_{opt}$ found a priori and utilizing $\kappa_{\!{max}}$ to optimize the maximum number of cluster-heads permitted per round.
  \item Adapting the cluster-head selection probability $P_{\!adp}$ according to existing network's state represented by $\kappa_{\!{max}}$ and number of alive SNs $\zeta$.
  \item Employing adaptive probability $P_{\!adp}$, jointly with $\kappa_{\!{max}}$ and energy factor $E$, to propose adaptive clustering algorithms.
  \item Employing the energy-distance ratio ($\digamma$) to assist SNs effectively select the best cluster-head to join, and to propose adaptive energy-efficient distance-based clustering algorithms jointly with $\kappa_{max}$, $E$, and $P_{\!adp}$.
  \item Verifying the performance of the proposed adaptive clustering algorithms, devised based on LEACH and Stable Election Protocol (SEP) algorithms, on randomly distributed heterogeneous SNs produced with different initial energy-levels.
\end{itemize}

\section{Background and Related Work}
\label{sec:background}

Clustering is a systematic practice in WSNs to facilitate routing and collecting data in a timely and energy-efficient manner~\cite{Rajesh2020, DAANOUNE2021, RCH-LEACH, Adnan2019, Dimple2019}. Clustering approaches are proposed to prolong the network lifetime, with their focus harnessed on devising cluster-head selection methods that evenly balance the energy consumption among SNs~\cite{sabor2017comprehensive, liang2019research, salam2020performance, alharbi2021towards, Adil2020}. To efficiently overcome the problem of energy consumption in clustering algorithms, LEACH protocol is proposed by Heinzelman et al.~\cite{Heinzelman} to enable self-organization of SNs and support cluster-head selection. In this protocol, a cluster-head is randomly rotated among all SNs to avoid fast energy depletion of a specific SN, so as to evenly distribute and balance the energy consumption among SNs and thus to prolong the network lifetime. The maximum number of clusters expected to minimize the total energy consumption of the SNs is calculated.

Similarly, Heinzelman et al.~\cite{Heinzelman2} evaluate the performance of LEACH algorithm with direct communication, Minimum Transmission-Energy (MTE), and static clustering algorithms. The effectiveness of LEACH algorithm outperforms the other algorithms by reducing the communication energy and increasing SN's lifetime. Nevertheless, LEACH algorithm suffers from numerous drawbacks in that it randomly selects cluster-heads at each round. SNs have equal probabilities to become cluster-heads regardless of their existing residual energy-levels and number of remaining SNs in the network, which in turn may choose a cluster-head with low energy-level.

Smaragdakis et al.~\cite{SEP_SANPA04} propose SEP protocol that selects cluster-heads such that time period to death of first SN is prolonged. SEP assumes SNs with different initial energy so as to tackle heterogeneity of the network. Improvements on SEP algorithm have been proposed to produce enhanced energy consumption and performance~\cite{SEP_2014}. Nurlan et al.~\cite{nurlan2021EZSEP} extends Z-SEP routing algorithm to propose an EZ-SEP that connects SNs to BS via a hybrid method. Though, some SNs communicate with the BS directly while some other SNs form together clusters to transfer data. Cluster-heads are selected based on residual energy of each SN.

Furthermore, Pranathi et al.~\cite{robust2020} present the centralized approach appeared in LEACH, in which a cluster-head is a single-point of failure that leads to information loss when it is transferred to BS. They propose a distributed, robust average-consensus routing protocol that makes SNs exchange information directly to BS to overcome the problem of centralization. Despite the robustness of their consensus algorithm in collecting and transmitting data, it suffers from a high energy consumption required by each SN to transmit its data directly to BS. As such, the proposed algorithm is modified to present an energy-efficient hybrid approach that mitigates the energy consumption to a level close to LEACH protocol. 

Several clustering protocols for lifetime maximization of WSNs are also proposed in the literature to tackle such clustering issues. Rawat et al.~\cite{rawat2021novel} present an energy-efficient LEACH-based routing protocol for network lifetime's enhancement. Elmasry et al.~\cite{EESRA} propose a scalable energy-efficient algorithm that adopts a three-layer hierarchy to minimize workload on cluster-heads and randomize the cluster-head selection procedure. Loganathan et al.~\cite{loganathan2020energyCentroid} propose an energy-efficient clustering algorithm that uses a distance centroid algorithm to design each cluster. The energy centroid of each cluster is computed and employed together with the energy threshold of each SN to choose a cluster-head. Muthukumaran et al.~\cite{K2018energy} present a hierarchical routing that uses cluster identification, multi-hop, and multi-level routing schemes to evaluate their effectiveness to produce energy-efficient clustering. Priyadarshi et al.~\cite{priyadarshi2019energy} propose a clustering algorithm that uses threshold functions and residual energy of each SN to select cluster-heads. Lin et al.~\cite{lin2021research} modify the threshold function of cluster-head selection to present a strategy for forming clusters of the network.

Qabouche et al.~\cite{qabouche2021hybrid} propose a static routing protocol that combines clustering and multi-hop routing. Data is transmitted in each round through independent SNs called gateways, which connect cluster-head SNs with the BS. Pour et al.~\cite{pour2021new} present an energy-aware cluster-head selection based on residual energy of a SN, centrality of a SN, and number of neighboring SNs. In addition, Chauhan et al.~\cite{chauhan2020mobile} propose a cluster-head selection strategy that splits the WSN into rectangular regions, each of which is managed by a cluster-head. A nature-inspired algorithm employs residual energy of each SN and distance from a SN to a sink node to select cluster-heads of the network. Bhola et al.~\cite{bhola2020genetic} adopts a genetic-based approach to optimally find route to destination nodes such that energy consumption is mitigated. The genetic algorithm utilizes its fitness function to optimize the network's performance, thus if the network performance is decreased then the genetic algorithm alters the route such that network's efficiency is increased. Wang et al.~\cite{Wang2020GA} also present a genetic algorithm that selects the best cluster-heads and combines them in a single chromosome to find the optimal routing path, with a fitness function that considers load balancing among SNs and minimum energy consumption.

In addition, Lata et al.~\cite{Lata2020} propose a reliable clustering algorithm that adopts a fuzzy-based clustering to formulate cluster, as well as select cluster heads and vice clusters, in a centralized-based manner such that network's lifetime is maximized and energy load is balanced. Ant-colony optimization is utilized by Liang et al.~\cite{liang2019research} to present an optimized multi-hop routing protocol, where energy consumption per round is employed to compute number of cluster-heads for an improved LEACH-based protocol. Shukry et al.~\cite{shukry2021stable} propose a Node Stable Routing protocol that stabilizes data transmission exchanged between SNs. The protocol characterizes a SN's stability by its residual energy and number of successive hops required to reach a destination SN. The source SN formalizes a stable path to destination such that energy consumption is mitigated.

A missing factor in proposed clustering strategies of WSNs is to actively account for network state at run-time~\cite{SuleimanP1_2019, SuleimanP2_2019, SuleimanP3_2020}. Such a state varies very often because SNs by the time suffer from gradual decrements in their residual energy, which in turn affects the network's configuration. Accordingly, cluster-heads are not adaptively elected based on existing states of the network and hence the total energy of network would not be fairly distributed among SNs leaving network's space proportionally covered. The clustering protocol in this paper employs an adaptive probabilistic function that formulates clusters so that QoS performance is enhanced. It computes a near-optimal distance between SNs such that energy consumption is reduced and thus network lifetime is maximized.

\section{Network Model}
\label{sec:NetMod}

The network model comprises $N$ heterogeneous SNs with similar computation and communication capabilities. Such SNs are randomly distributed on a space of $M$x$M$ area, with one BS positioned in the middle. The SNs and BS have fixed locations. However, all SNs are energy-constrained. Each SN starts with different initial amount of energy, has fixed Identifier (ID) address and position, and has enough power to reach the BS if a SN needs to act as a cluster-head. On the other side the BS works without energy constraint because it is assumed to have adequate energy supply; and as such the energy consumption of BS is not considered in the mathematical analysis and design evaluation. As in SEP protocol, SNs are divided into two types: normal and advanced SNs. Normal SNs start with low initial energy-level represented by $E_{0}$. Advanced SNs start with high initial energy-level represented by $\nolinebreak{E_0(1\!+\!a)}$, where $a$ is constant. Percentage of SNs that are advanced is $m$. Table~I summarizes notations used throughout the paper.
\begin{table*}[!t]
\label{tab:notations}
\begin{tabular}{c|l}
\hline
Notation              & Definition                                                                             \\ \hline
$a$                   & Heterogeneity factor for advanced SNs                                                  \\
$\alpha$              & Energy constant                                                                        \\
$\beta$               & Distance constant                                                                      \\
$d$                   & Distance between two adjacent SNs                                                      \\
$d_0$                 & Distance threshold                                                                     \\
$d_{B\!S}$            & Distance between a SN and a BS                                                         \\
$d_{C\!H}$            & Distance between a SN and a cluster-head                                               \\
$d_{opt}$             & Optimal distance between a cluster-head and its cluster-members                        \\
$\epsilon_{\!{f\!s}}$ & Energy dissipated by transmitter to run power amplifier using free space ($f\!s$) model \\
$\epsilon_{\!{m\!p}}$ & Energy dissipated by transmitter to run power amplifier using multipath fading ($m\!p$) model \\
$E_0$                 & Initial energy-level of normal SNs                                                     \\
$E_{init}$            & Initial energy of a SN                                                                 \\
$E_{res}$             & Current residual energy of a SN                                                        \\
$E_{\bar{e}}$         & Energy dissipated per bit to run the transmitter and receiver radio electronics        \\
$E_{\!{T\!X}}(l,d)$   & Energy dissipated to transmit $l$ bits through a distance $d$                          \\
$E_{\!{R\!X}}(l)$     & Energy dissipated to receive $l$ bits                                                  \\
$E_{\!{D\!A}}$        & Energy dissipated to aggregate $l$ bits                                                \\
$E_{\!{member}}$      & Energy dissipated by a cluster-member                                                  \\
$E_{{C\!H}}$          & Energy dissipated by a cluster-head                                                    \\
$E_{{cluster}}$       & Energy dissipated in each cluster                                                      \\
$E_{{Total}}$         & Energy dissipated by all clusters                                                      \\
$\digamma$            & Energy-distance ratio                                                                  \\
$\mathbb{G}$          & Set of non-elected SNs                                                                 \\
$\kappa$              & Number of clusters per round                                                           \\
$\kappa_{max}$        & Maximum number of cluster-heads permitted per round                                    \\
$l$                   & Size of data packet in bits                                                            \\
$m$                   & Percentage of advanced SNs distributed in the field                                    \\
$M$                   & Field's dimension                                                                      \\
$N$                   & Number of SNs distributed in the field                                                 \\
$p$                   & Probability of selecting a cluster-head                                                \\
$P_{\!adp}$           & Adaptive probability                                                                   \\
$r$                   & Current round number                                                                   \\
$s$                   & A selected SN from the set $\mathbb{G}$                                                \\
$T(s)$                & Threshold for selecting a SN as a cluster-head                                         \\
$T_{E}(s)$            & Threshold for selecting a SN as a cluster-head based on energy                         \\
$\zeta$               & Number of alive SNs in the current round                                               \\ \hline
\end{tabular}
\caption{Summary of Notations}
\end{table*}

All cluster-members sense the environment at a fixed rate and always have data to send to their corresponding cluster-heads. Each cluster has one cluster-head, and each cluster-member is allocated to one cluster-head, i.e., fuzzy membership is not allowed. Cluster-members communicate with the BS via their corresponding cluster-head by means of symmetric communication channel, i.e., same energy is required to transmit data packets from source to destination SNs and vise versa for a given Signal-to-Noise Ratio (SNR). Each cluster-head aggregates and then transmits data directly to the BS, and the BS receives compressed data.

The size of the transmitted and received data packet is fixed, $l$ bits. Each SN consumes energy when it receives ($E_{\!{R\!X}}$), aggregates ($E_{\!{D\!A}}$), and transmits ($E_{\!{T\!X}}$) data packets of size $l$ bits. Such energy consumption only depends on a distance $d$ between source and destination SNs.
For the distance between any two adjacent SNs, the free space model $d^2$ is employed if the distance between source and destination SNs is smaller than a threshold $d_0$ ($d\!\leq\!d_0$), otherwise ($d\!>\!d_0$) the multi-path fading model $d^4$ is employed. The network simulation parameters are shown in Table~II, which illustrates the maximum number of rounds, field dimensions in meters, SN parameters, values for SNs' heterogeneity, and parameters of the energy model.
\begin{table*}[!h]
\label{tab:setup}
\centering
\begin{tabular}{lc}
\hline
\multicolumn{2}{l}{\textbf{Field Dimension - Maximum $x$ and $y$ (in meters)}}                 \\
Maximum $x_m$ and $y_m$ coordinates in the field   & $100\text{ x }100 m^{2}$                \\
$x$ and $y$ coordinates of the sink node ($0.5x_m\text{ x }0.5y_m$)  & $50\text{ x }50$      \\  \hline
\multicolumn{2}{l}{\textbf{SN Parameters}}                                                   \\
Maximum number of rounds                    & 3000                                           \\
Number of SNs in the field ($N$)            & 100                                            \\
Initial optimal election probability of a SN to become cluster-head ($p$)   & 0.1            \\ \hline
\multicolumn{2}{l}{\textbf{Values for SNs Heterogeneity}}                                    \\
Percentage of advanced SNs in the field ($m$)  & 0.1                                         \\
Heterogeneity factor ($a$) for advanced SNs    & 1                                           \\ \hline
\multicolumn{2}{l}{\textbf{Energy Model (All Values in Joules)}}                             \\
Initial energy ($E_0$)               & $0.5~J/bit$                                 \\
Transmitting ($E_{{T\!X}}$) and receiving ($E_{\!{R\!X}}$) energy & $0.5\text{ x }10^{−9}~J/bit$ \\
Data aggregation energy ($E_{\!{D\!A}}$)    & $5\text{x}10^{−9}~J/bit$                                \\
Free-space model transmit amplifier energy ($\mathcal{\epsilon}_{\!{f\!s}}$)  & $10\text{ x }10^{−12}~J/bit/m^{2}$ \\
Multi-path model transmit amplifier energy ($\mathcal{\epsilon}_{{m\!p}}$) & $0.0013\text{ x }10^{−12}~J/bit/m^{4}$ \\ \hline
\end{tabular}
\caption{Simulation Parameters}
\captionsetup{justification=centering}
\end{table*}

\section{Mathematical Model}
\label{sec:MathMod}

A SN typically consumes energy when it receives, aggregates, and transmits data packets. The mathematical model adopts the simple energy dissipation model explained in~\cite{Heinzelman}. A transmitter SN consumes energy to run the radio electronics and power amplifier, and a receiver SN consumes energy to only run the radio electronics.
\begin{equation}
\label{equ:E_TX}
E_{\!{T\!X}} =
\begin{cases}
     lE_{\bar{e}} + l\epsilon_{\!{f\!s}} d^2, \;\;\;\;\;\;\;\;\;\;  d \leq d_0  \\
     lE_{\bar{e}} + l\epsilon_{\!{m\!p}} d^4, \;\;\;\;\;\;\;\;\;  d > d_0
\end{cases}
\end{equation}
\begin{equation}
\label{equ:E_RX}
E_{\!{R\!X}} = lE_{\bar{e}}
\end{equation}

The $E_{\!{T\!X}}(l,d)$ is the energy dissipated to transmit $l$ bits through a distance $d$, whereas $E_{\!{R\!X}}(l)$ is the energy dissipated to receive $l$ bits. The $E_{\bar{e}}$ is the energy dissipated per bit to run the transmitter and receiver radio electronics. Also, $\epsilon_{\!{f\!s}}$ and $\epsilon_{\!{m\!p}}$ are the energy dissipated by transmitter to run the power amplifier using the free space ($f\!s$) model and the multi-path fading ($m\!p$) model, respectively. However, the free space ($f\!s$) and multi-path fading ($m\!p$) models are used depending on the distance between the source and destination SNs. The distance is first compared with a threshold value, $\nolinebreak{d_0 = \sqrt{\frac{\epsilon_{\!{f\!s}}}{\epsilon_{\!{m\!p}}}}}$. The free space model ($d^2$) is employed if the distance $d$ is $d\!\leq\!d_0$; otherwise, the multi-path fading model ($d^4$) is employed. Mathematically, the free space model ($d^2$) is used in this paper for intra-cluster communication because the distance $d$ is assumed to be $d\!\leq\!d_0$, whereas the multi-path fading model ($d^4$) is utilized for communication between cluster-head and BS because the distance $d$ is assumed to be $d\!>\!d_0$.

\subsection{Formalizing $d_{opt}$ and $\kappa_{max}$}

It is assumed that $N$ SNs are randomly distributed over an area of $M$x$M$. Since the number of clusters per round is $\kappa$, then each cluster comprises $\frac{N}{\kappa}$ SNs. 
Each cluster-head dissipates energy $E_{{C\!H}}$ when it receives data from $\nolinebreak{(\frac{N}{\kappa}\!-\!1)}$ cluster-members (denoted by $E_{\!{R\!X}}$), as well as dissipates energy when it aggregates ($E_{\!{D\!A}}$) and transmits ($E_{{T\!X}}$) such data to the BS. 
\begin{equation}
\label{equ:E_CH}
\begin{split}
   E_{{C\!H}} & = E_{\!{R\!X}} + E_{\!{D\!A}} + E_{{T\!X}} \\
                        & = lE_{\bar{e}} (\frac{N}{\kappa}\!-\!1) + lE_{\!{D\!A}}\frac{N}{\kappa} + lE_{\bar{e}} + l\epsilon_{{m\!p}} d_{B\!S}^4
\end{split}
\end{equation}

where $d_{BS}$ is the distance between a cluster-head and BS. Each cluster-member dissipates energy $E_{\!{member}}$ when it transmits data to its cluster-head.
\begin{equation}
\label{equ:E_member}
   E_{\!{member}} = lE_{\bar{e}} + l\epsilon_{\!{f\!s}} d_{C\!H}^2
\end{equation}

where $d_{C\!H}$ is the distance from a cluster-member to its cluster-head. The energy dissipated in a cluster $E_{{cluster}}$ is equal to the energy dissipated by the cluster-head $E_{{C\!H}}$ and its cluster-members $(\frac{N}{\kappa}-1)E_{\!{member}}$.
\begin{equation}
\label{equ:E_Cluster}
\begin{split}
   E_{{Cluster}} & = E_{{C\!H}} + (\frac{N}{\kappa}\!-\!1)E_{\!{member}} \\
                           & \approx E_{{C\!H}} + (\frac{N}{\kappa})E_{\!{member}}
\end{split}
\end{equation}

Then, the overall energy consumption $E_{{Total}}$ dissipated by all clusters is
\begin{equation}
\label{equ:E_Total}
\begin{split}
   E_{{Total}} & = \kappa E_{{Cluster}} \\
                         & = \kappa E_{{C\!H}} + NE_{\!{member}} \\
                         & = \kappa \big( lE_{\bar{e}} (\frac{N}{\kappa}\!-\!1) + lE_{\!{D\!A}}\frac{N}{\kappa} + lE_{\bar{e}} + l\epsilon_{{m\!p}} d_{B\!S}^4 \big) \\
                         &\;\;\;\;\;\; + N(lE_{\bar{e}} + l\epsilon_{\!{f\!s}} d_{C\!H}^2)   \\
                         & = 2lE_{\bar{e}}N + lE_{\!{D\!A}} N + l\epsilon_{{m\!p}} d_{B\!S}^4 \kappa + N l\epsilon_{\!{f\!s}} d_{C\!H}^2
\end{split}
\end{equation}

The objective is to formulate a distance $d_{opt}$ between each cluster-head and its cluster-members, such that the overall energy consumption of the network is mitigated. The network field is partitioned into equal circles, each of which is of size $\nolinebreak{2\pi d_{\!{opt}}^2}$. The expected number of clusters per round is
\begin{equation}
\label{equ:K}
   \kappa = \frac{M^2}{2\pi d_{opt}^2}
\end{equation}

By replacing $\kappa$ with its new value, replacing $d_{C\!H}$ with $d_{opt}$, and deriving the overall energy consumption with respect to $d_{opt}$, then the final formulation becomes
\begin{equation}
\label{equ:E_Total_new}
   E_{{Total}} = 2lE_{\bar{e}}N + lE_{\!{D\!A}} N + l\epsilon_{{m\!p}} d_{B\!S}^4 \frac{M^2}{2\pi d_{opt}^2} + N l\epsilon_{\!{f\!s}} d_{opt}^2
\end{equation}
\begin{equation}
\label{equ:derive}
   \frac{\partial E_{{Total}}}{\partial d_{opt}} = 0
\end{equation}
\begin{equation}
\label{equ:dopt_new}
   d_{opt} = \sqrt[4]{\frac{\epsilon_{{m\!p}} M^2}{2\pi N\epsilon_{\!{f\!s}}}} d_{B\!S}
\end{equation}

This means that maximum number of clusters permitted per round to avoid any potential increase in the overall energy consumption of the network is
\begin{equation}
\label{equ:Kmax_new}
\begin{split}
   \kappa_{max} & = \frac{M^2}{2\pi d_{opt}^2} \\
                & = \sqrt{\frac{N\epsilon_{\!{f\!s}}}{2\pi \epsilon_{{m\!p}}}} \frac{M}{d_{B\!S}^2}
\end{split}
\end{equation}

The value of $\kappa_{max}$ achieved based on the distance-based clustering is compatible with the expected number of clusters calculated in~\cite{Heinzelman}.

\subsection{The impact of energy ($E$) and maximum number of clusters ($\kappa_{max}$)}

The objective is to first design an energy-efficient clustering algorithm that distributes the energy evenly among all SNs, such that no SN is frequently chosen to act as a cluster-head and runs out of energy before others. The $d_{opt}$ and $\kappa_{max}$ derived formerly are utilized to mitigate the overall energy consumption of the whole network. As explained in~\cite{Heinzelman}, all SNs have equal probabilities to become cluster-heads in each round. Each SN selects a random number between $[0,\!1]$ and elects itself to become a cluster-head in the current round if the random number is less than a threshold $T(s)$.
\begin{equation}
\label{equ:Ts}
T(s) =
\begin{cases}
     \frac{p}{1-p(r\!\!\!\!\mod\frac{1}{p})}             \;\;\;\;,               \;\;\; \text{if}~s\in \mathbb{G} \\
     \;\;\;\;\;\;\;\;\; 0 \;\;\;\;\;\;\;\;\;\;\; \;\;\;,               \;\;\; \text{otherwise}
\end{cases}
\end{equation}
where $r$ is the current round number, $s$ is the selected SN, $p$ is the probability of selecting a cluster-head, and $\mathbb{G}$ is the set of non-elected SNs. Accordingly, a SN that has not previously been a cluster-head is assumed to have more residual energy than others and qualified to become a cluster-head in next upcoming rounds. In LEACH algorithm, the overall energy is calculated by multiplying the average amount of energy of all SNs in each cluster by the total number $N$ of SNs.

This is accomplished by modifying the cluster-head selection probability to become a function $T_{E}(s)$ of current SN's energy relative to its initial energy, so that more probabilistic weight is given to SNs that hold high residual energy to become cluster-heads than others.
\begin{equation}
\label{equ:Ts_new}
T_{E}(s) =
\begin{cases}
     \frac{p}{1-p(r\!\!\!\!\mod\frac{1}{p})} \frac{E_{res}}{E_{init}} \;\;, \;\;\; \text{if}~s\in \mathbb{G}  \\
     \;\;\;\;\;\;\;\;\; 0 \;\;\;\;\;\;\;\;\;\;\;\;\;\;\;\; \;\;\;,              \;\;\;\; \text{otherwise}
\end{cases}
\end{equation}
where $E_{res}$ represents SN's current residual energy and $E_{init}$ represents SN's initial energy. In this case, SNs with high residual energy are more qualified and have high probabilities to become cluster-heads in the upcoming round. As such, the number of cluster-heads permitted per round $\kappa_{max}$ developed based on $d_{opt}$ are both employed to devise LEACH-based and SEP-based algorithms.

The number of clusters permitted per round is limited to $\kappa_{max}$ so that qualified SNs are evenly distributed throughout the network operation. However, $\kappa_{max}$ combined with $E$ manage and distribute the energy consumption evenly during the network operation instead of having several clusters in such a round. Tradeoff is obvious in applying $\kappa_{max}$ which may sometimes negatively overload some cluster-heads with extra data traffic and quickly drain their energy, especially when $\kappa_{max}$ is small and there are many SNs joining the selected cluster-heads. In this paper, the effect of $\kappa_{max}$ appears when $P_{adp}$ is computed as explained in the next section. 

\subsection{The impact of adaptive probability ($P_{\!adp}$)}

In LEACH and SEP algorithms, the probability of a SN to become a cluster-head is fixed. Such a probability however becomes inappropriate by the time because of increasing the number of dead SNs; which leads to increase the total number of rounds and decrease the number of elected cluster-heads per round. The probability of selecting a SN as a cluster-head should be adaptive and changed by the time depending on the network's state represented by existing network conditions and structures.

In this paper, the probability of selecting a SN to become a cluster-head is adapted at the end of each round. The $P_{\!adp}$ depends on the value of $\kappa_{max}$ permitted per round (calculated a priori) and the number of remaining (alive) SNs $\zeta$ in the network. Such a selection scheme in turn increases the probability of electing a SN from the remaining ones to become a cluster-head when the number of alive SNs $\zeta$ in the network is decreased.
\begin{equation}
\label{equ:Padp}
   P_{adp} = \frac{\kappa_{max}}{\zeta}
\end{equation}

Thus, LEACH and SEP algorithms are modified to have LEACH-$\kappa\!P$, LEACH-$\kappa E\!P$, SEP-$\kappa\!P$, and SEP-$\kappa E\!P$ algorithms. In LEACH-$\kappa\!P$ algorithm, the probability $P$ is adapted at the end of each round and the number of clusters permitted per round is limited to $\kappa_{max}$ to independently study the effect of $P_{adp}$. In LEACH-$\kappa E\!P$ algorithm, the probability $P$ is adapted at the end of each round, the number of clusters permitted per round is limited to $\kappa_{max}$, and the $E$ factor is employed in the cluster-head selection probability function. The same thing is applied for SEP-$\kappa\!P$ and SEP-$\kappa E\!P$ algorithms, respectively.

\subsection{The impact of the energy-distance ratio ($\digamma$)}

In LEACH and SEP algorithms, a SN joins a cluster-head based on the distance between them. The SN first computes the distance between it and all available cluster-heads, and accordingly the SN joins the nearest cluster-head. However, the difficulty of this practice is that a SN may join a cluster-head that does not have enough energy to serve all its cluster-members. Such a routine drains the energy of the selected cluster-head quickly and as a result leaves some regions uncovered. Though, as explained in~\cite{Jiao_2012}, a SN selects a cluster-head based on the energy-distance ratio $\digamma$, where a SN joins the cluster-head with the highest $\digamma$. The utilization of $\digamma$ ratio is more energy-efficient than joining a cluster-head based on distance only.
\begin{equation}
\label{equ:F}
   \digamma = \frac{E_{res}^{\alpha}}{d_{C\!H}^{\beta}}
\end{equation}

The $\kappa_{max}$, $E$, $P_{\!adp}$, and $\digamma$ are employed. Values of $\alpha$ and $\beta$ are varied, chosen to be $\nolinebreak{\alpha\!=\!\beta\!=\!1}$ to get LEACH-$\kappa E\!\digamma$\!-1-1, and then $P_{adp}$ is employed to get LEACH-$\kappa E\!\digamma$\!-1-1-$P$. Also, values of $\alpha$ and $\beta$ are chosen to be $\nolinebreak{\alpha\!=\!1}$ and $\nolinebreak{\beta\!=\!2}$ to get LEACH-$\kappa E\!\digamma$\!-1-2, and then $P_{adp}$ is employed to get LEACH-$\kappa E\!\digamma$\!-1-2-$P$. Same thing is applied for the proposed SEP algorithms to get SEP-$\kappa E\!\digamma$\!-1-1, SEP-$\kappa E\!\digamma$\!-1-1-$P$, SEP-$\kappa E\!\digamma$\!-1-2, and SEP-$\kappa E\!\digamma$\!-1-2-$P$, respectively. For figures of the number of dead SN, the round number period of comparison is chosen to be between $\nolinebreak{800\!-\!2000}$, where the major difference between proposed algorithms appears. For figures of the remaining energy, the round number period of comparison is chosen to be between $\nolinebreak{1000\!-\!1600}$.

\section{Evaluation}
\label{sec:eval}

The adaptive-based clustering algorithms are developed by utilizing LEACH and SEP. The QoS performance impact of $P_{adp}$ on the probability of dead SNs and the energy consumption are examined. The $\kappa_{max}$, $E$, and $\digamma$ performance parameters are employed and evaluated against each others.

\subsection{The impact of $P_{adp}$ on death rate of SNs by utilizing $\kappa_{max}$, $E$, and $\digamma$}

\begin{figure}[!t]
\centering
\captionsetup{justification=centering}
	  \includegraphics[width=0.65\textwidth]{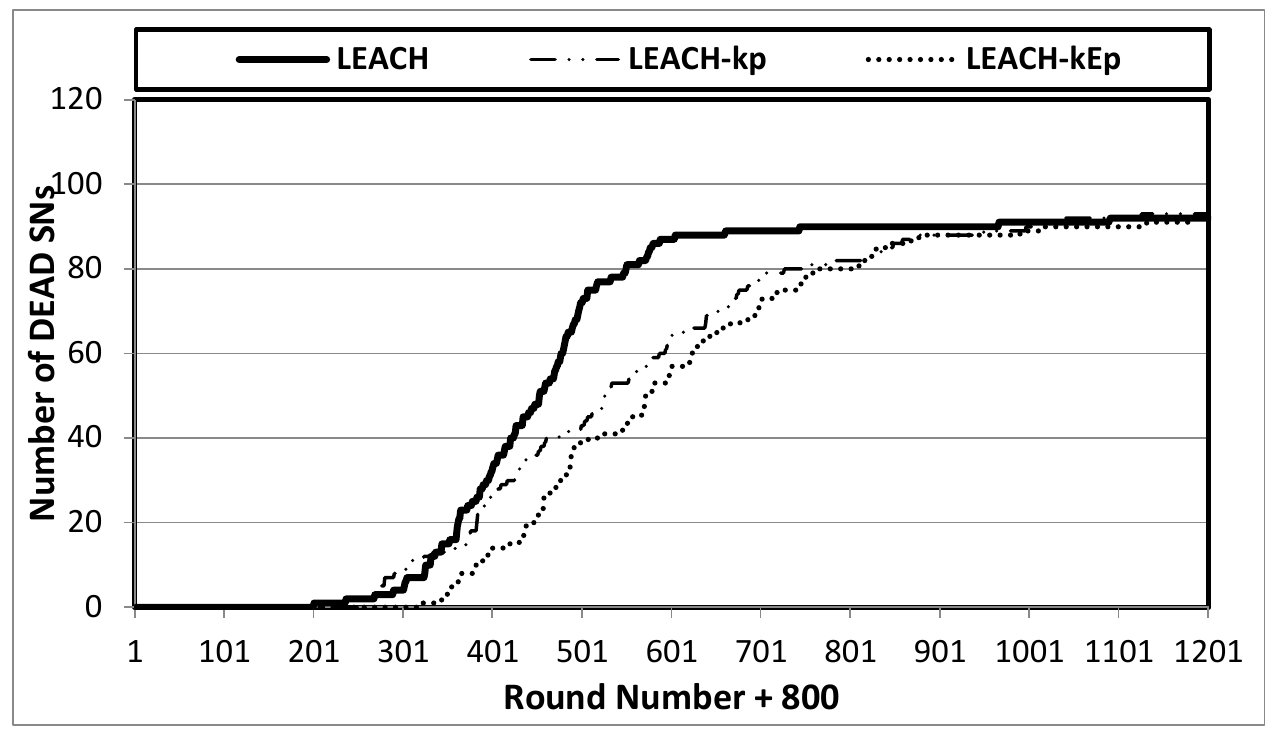}
	  \caption{\small{The Effect of $\kappa_{max}$, $E$, and $P_{adp}$ on Dead SNs per Round in LEACH-based Algorithms}}
      \label{fig:LEACH_Comparison_Dead}
\end{figure}

The QoS performance of the proposed LEACH algorithms is evaluated by computing the likelihood of dead SNs as shown in Figure~\ref{fig:LEACH_Comparison_Dead}. The impacts of $\kappa_{max}$, $E$, and $P_{adp}$ performance factors are examined and analyzed. Commonly, $\kappa_{max}$ maintains a stable QoS performance throughout the network operation due to the limited number of maximum cluster-heads permitted per round. In this condition, $\kappa_{max}$ puts more constraints and controls on electing cluster-heads per round by making it optimal, which preserves and defers some qualified SNs to become cluster-heads in upcoming rounds and accordingly distributes the energy evenly among all SNs throughout the network operation. Consequently, $\kappa_{max}$ positively influences the probability of dead SNs per round. That is because qualified SNs are evenly distributed throughout the network lifetime, instead of having most of them running early as cluster-heads which in turn would deplete their energy rapidly. The effect of $\kappa_{max}$ appears clearly when $P_{adp}$ is adjusted at the end of each round, because $P_{adp}$ accounts for $\kappa_{max}$ and number of alive SNs $\zeta$ exist in the network.

Examining the impact of utilizing $P_{adp}$, the formulation of LEACH-$\kappa\!P$ algorithm developed from employing $P_{adp}$ improves the QoS performance of LEACH algorithm. That is because the cluster-head probability is adapted based on existing network states. Since $\kappa_{max}$ is fixed throughout the time of network operation, $P_{adp}$ increases when number of alive SNs $\zeta$ decreases. Increasing $P_{adp}$, however, puts less constraint on selecting a SN to become a cluster-head, and hence increases the likelihood of qualified SNs to act as cluster-heads. The formulation of LEACH-$\kappa\!P$ algorithm increases the stability period and decreases the probability of dead SNs, as compared to LEACH algorithm.

Exploiting $E$ factor to devise LEACH-$\kappa E\!P$ algorithm improves the stability period and decreases the probability of dead SNs, as compared to LEACH and LEACH-$\kappa\!P$ algorithms. The probability function of LEACH-$\kappa E\!P$ algorithm employs the current SNs' residual energy to thus offer more probabilistic weight to SNs with high energy to become cluster-heads than others. Such energy considerations would in turn distribute the energy evenly among all SNs throughout the network lifetime. Overall, engaging $E$ and/or $P_{adp}$ to LEACH algorithm improves the stability period and mitigates the likelihood of dead SNs.
\begin{figure}[!t]
\centering
\captionsetup{justification=centering}
	  \includegraphics[width=0.65\textwidth]{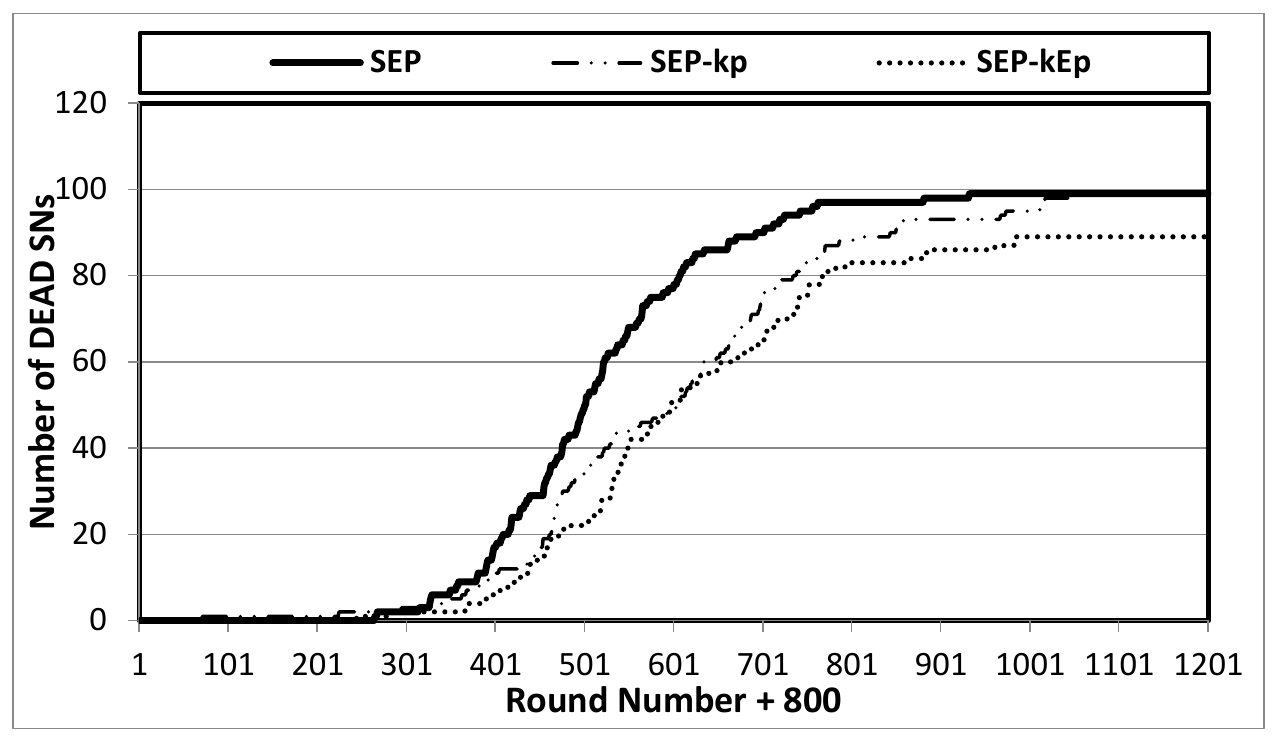}
	  \caption{\small{The Effect of $\kappa_{max}$, $E$, and $P_{adp}$ on Dead SNs per Round in SEP-based Algorithms}}
      \label{fig:SEP_Comparison_Dead}
\end{figure}

The same thing is applied when SEP algorithms are assessed as shown in Figure~\ref{fig:SEP_Comparison_Dead}. SEP-$k\!P$ algorithm surpasses the performance of SEP algorithm by enhancing the probability of dead SNs and stability period throughout most of the network lifetime. In contrast, SEP-$\kappa E\!P$ algorithm has the greatest stability period in comparison with SEP and SEP-$\kappa\!P$ algorithms. As well, SEP-$\kappa E\!P$ algorithm improves the performance of SEP algorithms by decreasing the probability of dead SNs, due to combining the effect of $E$ and $P_{adp}$ performance metrics.

Generally, SEP, SEP-$\kappa\!P$, and SEP-$\kappa E\!P$ algorithms outperform LEACH, LEACH-$\kappa\!P$, and LEACH-$\kappa E\!P$ algorithms, respectively. SEP algorithm basically has different probabilities for advanced and normal SNs; that is $P_{adv}$ for advanced SNs and $P_{nrm}$ for normal SNs. SEP algorithm offers more weight to advanced SNs (by a factor of $\nolinebreak{(1\!+\!a)}$) in the original probability function to become cluster-heads than others due to their high residual energy-levels, as compared to LEACH algorithm which allocates equal probabilities to all SNs. Overall, utilizing the effect of $P_{adp}$ and $E$ performance factors together on LEACH and SEP algorithms improves the stability period and mitigates the likelihood of dead SNs per round.
\begin{figure}[!h]
\centering
\captionsetup{justification=centering}
	  \includegraphics[width=0.65\textwidth]{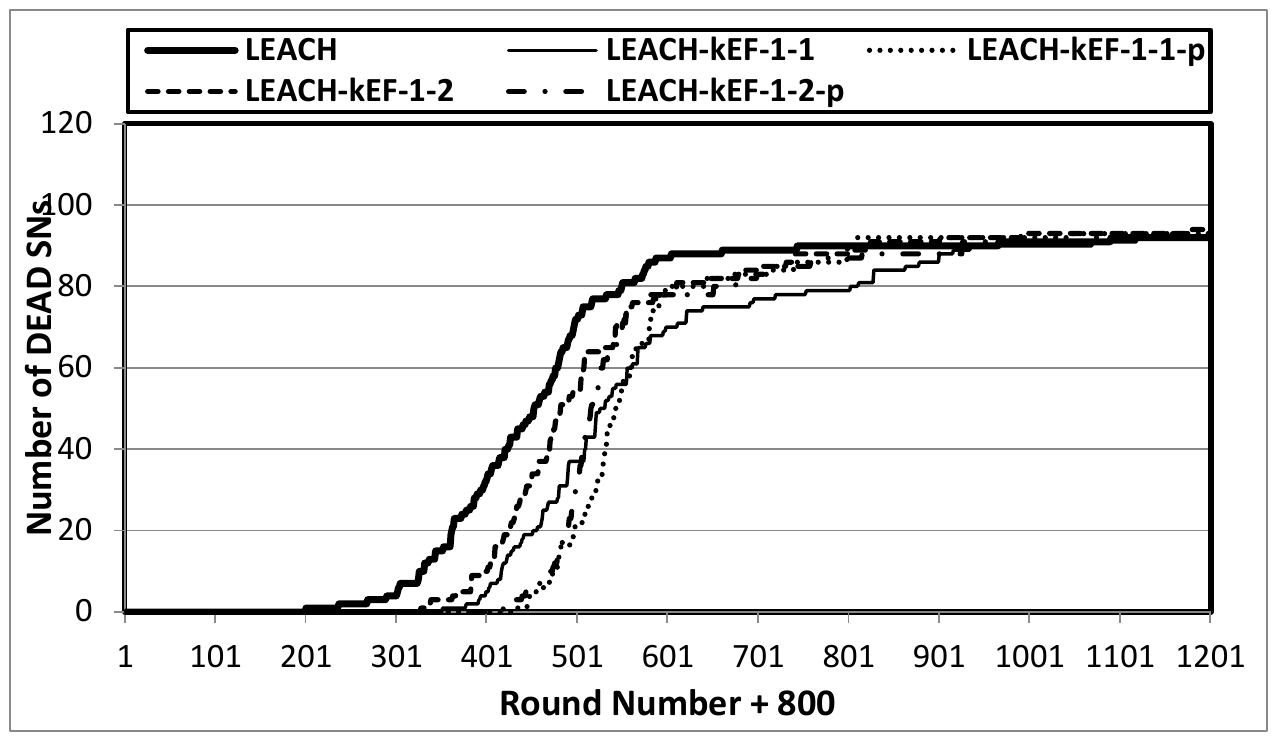}
	  \caption{\small{The Effect of $P_{adp}$ and $\digamma$ on Dead SNs per Round in LEACH-based Algorithms}}
      \label{fig:LEACH_kEFpab_Comparison_Dead}
\end{figure}
\begin{figure}[!h]
\centering
\captionsetup{justification=centering}
	  \includegraphics[width=0.65\textwidth]{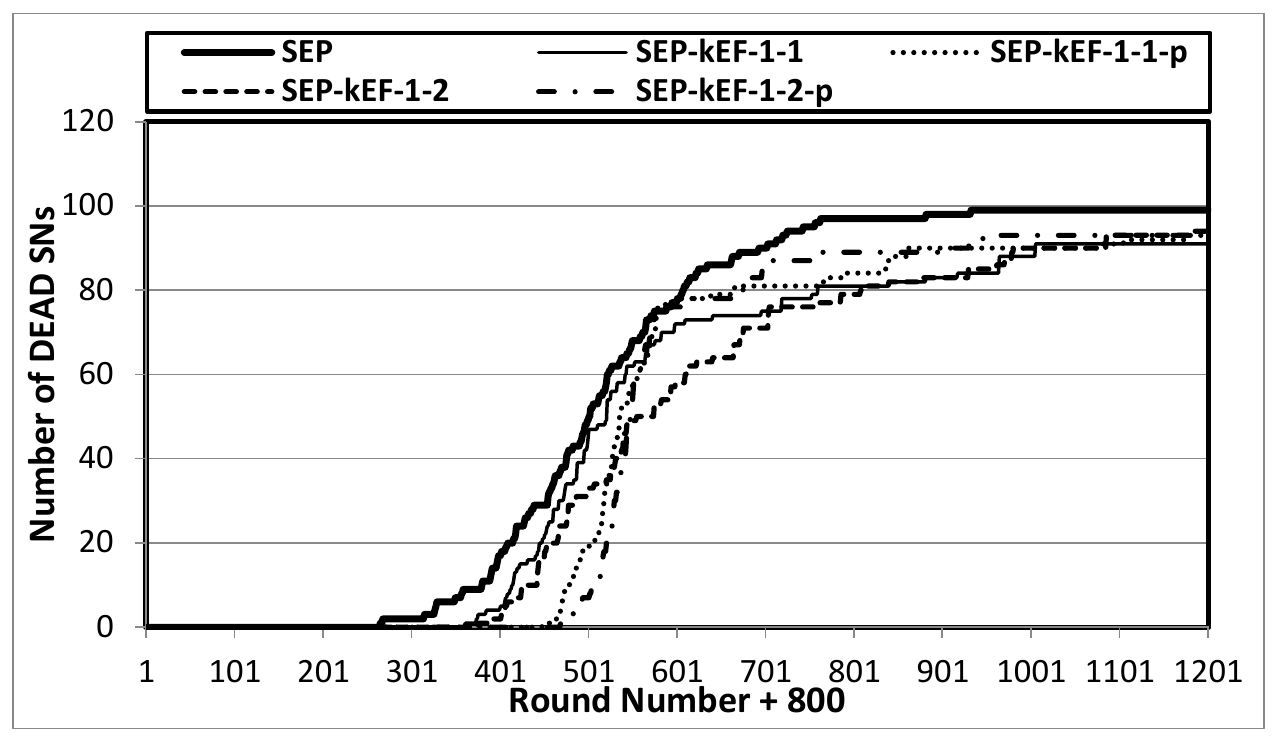}
	  \caption{\small{The Effect of $P_{adp}$ and $\digamma$ on Dead SNs per Round in SEP-based Algorithms}}
      \label{fig:SEP_kEFpab_Comparison_Dead}
\end{figure}
The QoS performance of the proposed LEACH algorithms presented in Figure~\ref{fig:LEACH_kEFpab_Comparison_Dead} is evaluated on the likelihood of dead SNs by considering the impact of $\kappa_{max}$, $E$, $P_{adp}$, and $\digamma$. It is obvious that employing $\digamma$ in combination with $\kappa_{max}$, $E$, and $P_{adp}$ improves the QoS performance of LEACH algorithm - apparently the stability period is increased and probability of dead SNs is decreased. As explained previously, SNs in LEACH algorithm join the nearest cluster-head. However, the nearest cluster-head does not necessarily have adequate energy to serve all its cluster-members for long time. Employing $\digamma$ ratio improves the QoS performance because it also utilizes the residual energy of cluster-heads. In this case, a SN joins a cluster-head with the highest $\digamma$ ratio. $P_{adp}$ combined with $\digamma$ improves the stability period and probability of dead SNs. The same thing is applied when SEP algorithms in Figure~\ref{fig:SEP_kEFpab_Comparison_Dead} are compared together.

\subsection{The impact of $P_{adp}$ on death rate of advanced SNs by utilizing $\kappa_{max}$, $E$, and $\digamma$}

\begin{figure}[!th]
\centering
\captionsetup{justification=centering}
	  \includegraphics[width=0.65\textwidth]{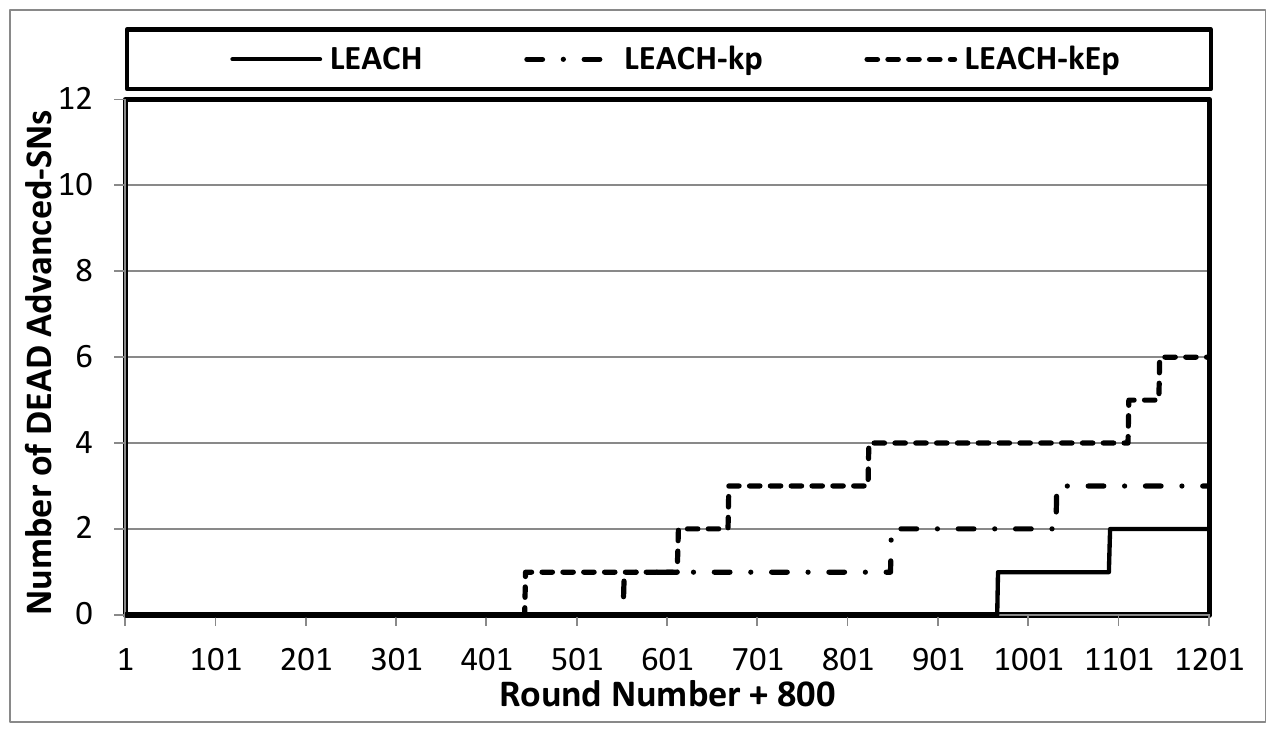}
	  \caption{\small{The Effect of $\kappa_{max}$, $E$, and $P_{adp}$ on Advanced Dead SNs per Round in LEACH-based Algorithms}}
      \label{fig:LEACH_Comparison_Advanced}
\end{figure}

The probability of advanced dead-SNs of the proposed LEACH-based algorithms is evaluated in Figure~\ref{fig:LEACH_Comparison_Advanced} by considering the effect of $\kappa_{max}$, $E$, and $P_{adp}$. Employing such performance measures on LEACH algorithm gives advanced SNs high probabilities to become cluster-heads and subsequently depletes their energy rapidly. LEACH algorithm by itself, however, selects cluster-heads randomly and does not guarantee to select the most energy-efficient SNs to become cluster-heads.
\begin{figure}[!th]
\centering
\captionsetup{justification=centering}
	  \includegraphics[width=0.65\textwidth]{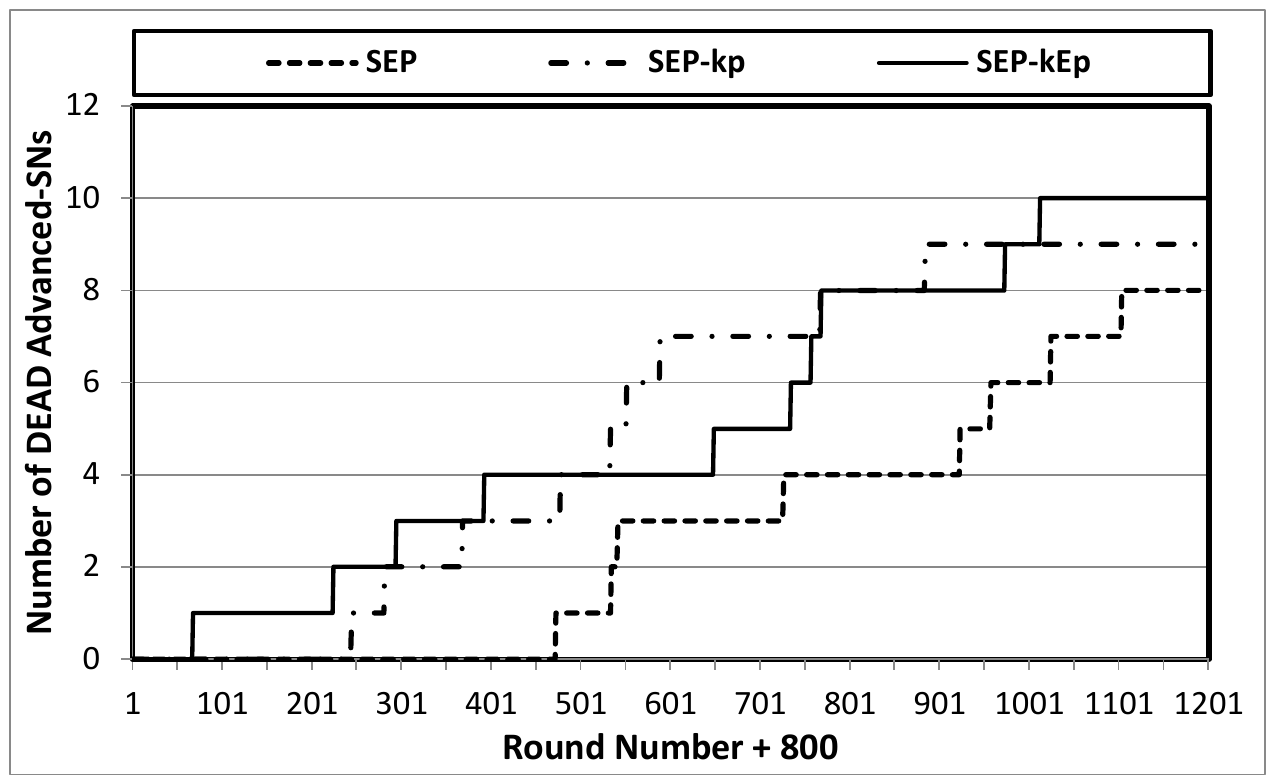}
	  \caption{\small{The Effect of $\kappa_{max}$, $E$, and $P_{adp}$ on Advanced Dead SNs per Round in SEP-based Algorithms}}
      \label{fig:SEP_Comparison_Advanced}
\end{figure}

In contrast, Figure~\ref{fig:SEP_Comparison_Advanced} compares the QoS performance of the proposed SEP-based algorithms on the likelihood of advanced dead-SNs by considering the effect of $\kappa_{max}$, $E$, and $P_{adp}$. Though, SEP algorithm utilizes high residual energy-levels of advanced SNs to accordingly elect them as cluster-heads early during the network lifetime; which thus decreases the number of advanced SNs throughout the network operation.

However, employing $P_{adp}$ increases the probability of electing advanced SNs and qualifies them more to become cluster-heads in upcoming rounds, especially that advanced SNs have high residual energy-levels as compared to normal SNs. This situation is obvious when SEP-$\kappa\!P$ algorithm is compared to SEP algorithm, in which $P_{adp}$ factor improves QoS performance of SEP algorithm. The $P_{adp}$ combined with $E$ formulates SEP-$\kappa E\!P$ algorithm, which in turn improves the performance and stability period, as well as has more impact on enhancing the likelihood of death rate for advanced SNs.

When SEP and LEACH algorithms are compared together, SEP algorithm offers high probabilistic weights to advanced SNs because of their high residual energy as compared to normal SNs. Such weights qualify advanced SNs by increasing their probabilities of being elected as cluster-heads and then die early. Most of advanced SNs are dead in case of SEP algorithm, whereas LEACH algorithm randomly selects cluster-heads without considering SN's residual energy which in turn may leave some of advanced SNs not being elected very often to act as cluster-heads despite their high residual energy-levels.

\begin{figure}[!h]
\centering
\captionsetup{justification=centering}
	  \includegraphics[width=0.65\textwidth]{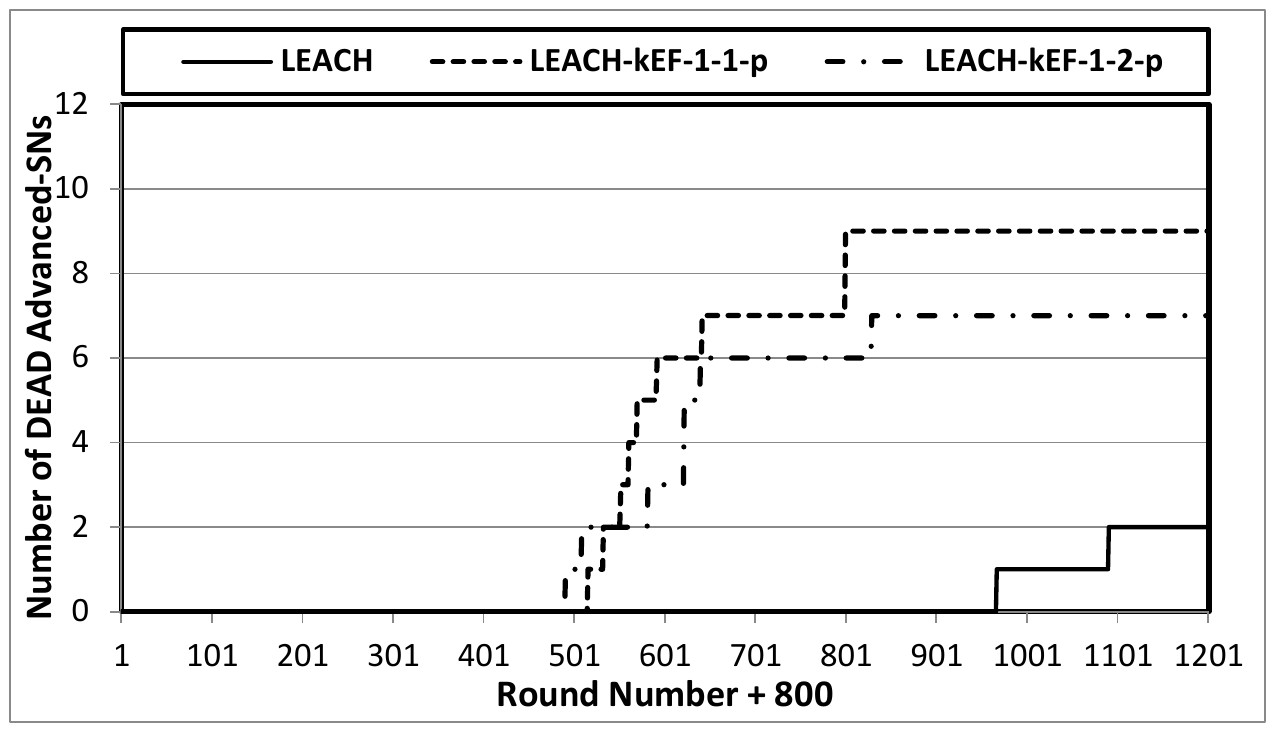}
	  \caption{\small{The Effect of $P_{adp}$ and $\digamma$ on Advanced Dead SNs per Round in LEACH-based Algorithms}}
      \label{fig:LEACH_kEFpab_Comparison_Advanced}
\end{figure}

Furthermore, the performance impact of $P_{adp}$ together with $\digamma$ ratio for the proposed LEACH-based algorithms in Figure~\ref{fig:LEACH_kEFpab_Comparison_Advanced} is compared on the likelihood of advanced dead-SNs. Advanced SNs die rapidly because of considering $E$ as a main factor in the modified LEACH-based algorithms to decide on cluster-heads to be elected in each round. Holding $\nolinebreak{\alpha\!=\!\beta\!=\!1}$ has more impact on advanced dead SNs than it is in the case of $\nolinebreak{\alpha\!=\!1}$ and $\nolinebreak{\beta\!=\!2}$; and the impact increases when $P_{adp}$ is employed.

\begin{figure}[!h]
\centering
\captionsetup{justification=centering}
	  \includegraphics[width=0.64\textwidth]{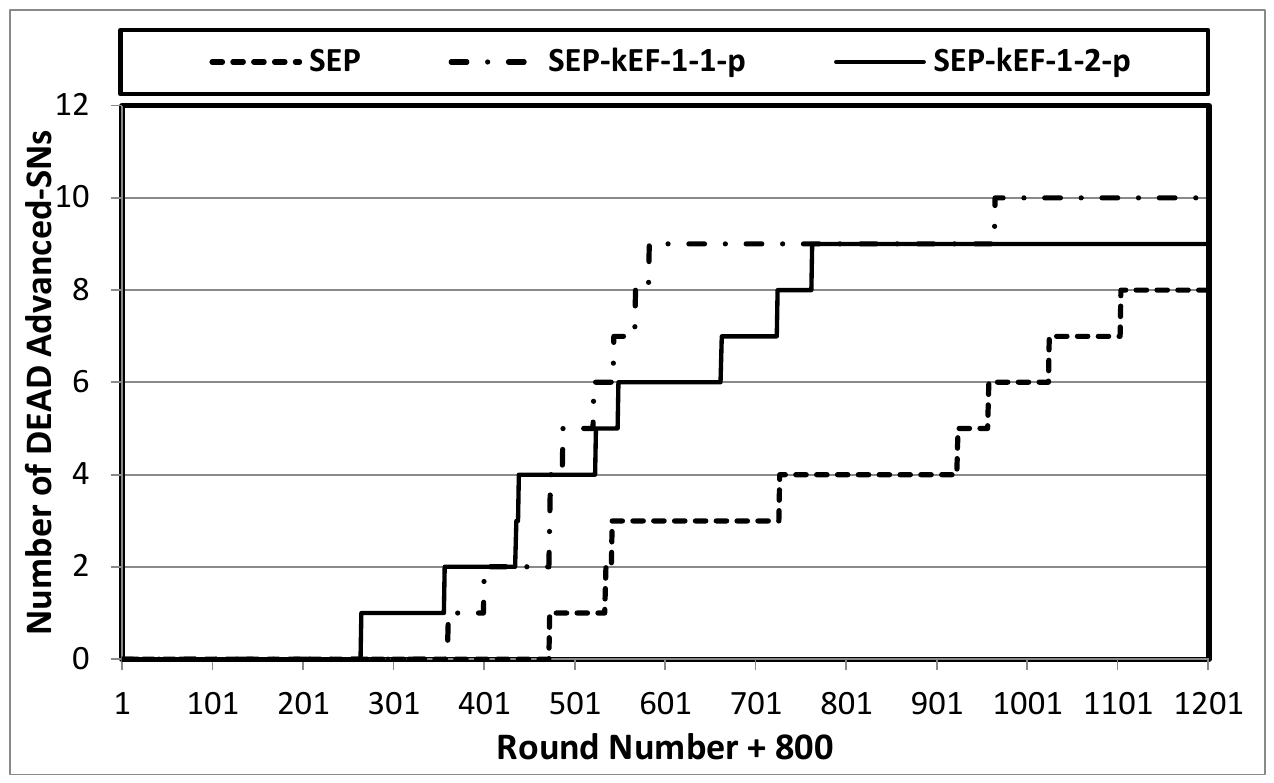}
	  \caption{\small{The Effect of $P_{adp}$ and $\digamma$ on Advanced Dead SNs per Round in SEP-based Algorithms}}
      \label{fig:SEP_kEFpab_Comparison_Advanced}
\end{figure}

In contrast, Figure~\ref{fig:SEP_kEFpab_Comparison_Advanced} assesses the performance of the proposed SEP-based algorithms on the number of advanced dead SNs by considering the effect of $P_{adp}$ and $\digamma$. The proposed SEP-based algorithms improve the QoS performance of SEP algorithm by enhancing the stability period and number of advanced dead-SNs. The reason is that the impact of energy and distance, instead of distance only, are considered at each SN to decide on a potential cluster-head to join in each round.

In addition, a cluster-head's energy mostly decreases by increasing the number of cluster-members by the time, by that a cluster-head consumes energy ($E_{R\!X}$ and $E_{D\!A}$) when it accepts a cluster-member. When a distance-based procedure is employed, a cluster-head may keep accepting many SNs because such SNs join the nearest cluster-head, which in turn may rapidly deplete cluster-head's energy. However the procedure of utilizing $\digamma$ ratio makes a SN first look at the cluster-head's energy relative to the distance between them, and then the SN joins the cluster-head with the highest $\digamma$. The latter procedure wisely distributes SNs evenly among existing cluster-heads, each of which with its capacity represented by its residual energy, and hence positively affects the overall QoS performance.
\begin{figure}[!t]
\centering
\captionsetup{justification=centering}
	  \includegraphics[width=0.65\textwidth]{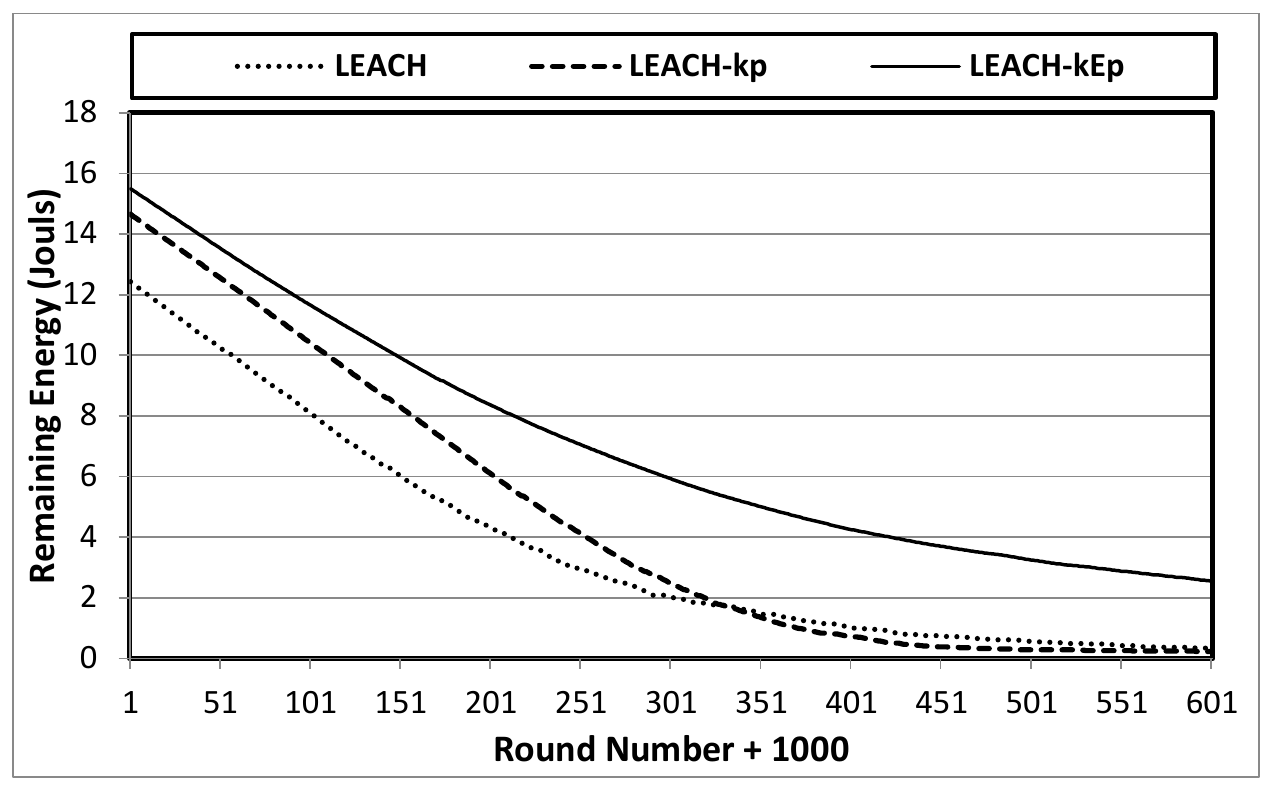}
	  \caption{\small{The Effect of $\kappa_{max}$, $E$, and $P_{adp}$ on Energy Consumption per Round in LEACH-based Algorithms}}
      \label{fig:LEACH_Comparison_remEnergyDead}
\end{figure}

\subsection{The impact of $P_{adp}$ on energy consumption}

\begin{figure}[!h]
\centering
\captionsetup{justification=centering}
	  \includegraphics[width=0.65\textwidth]{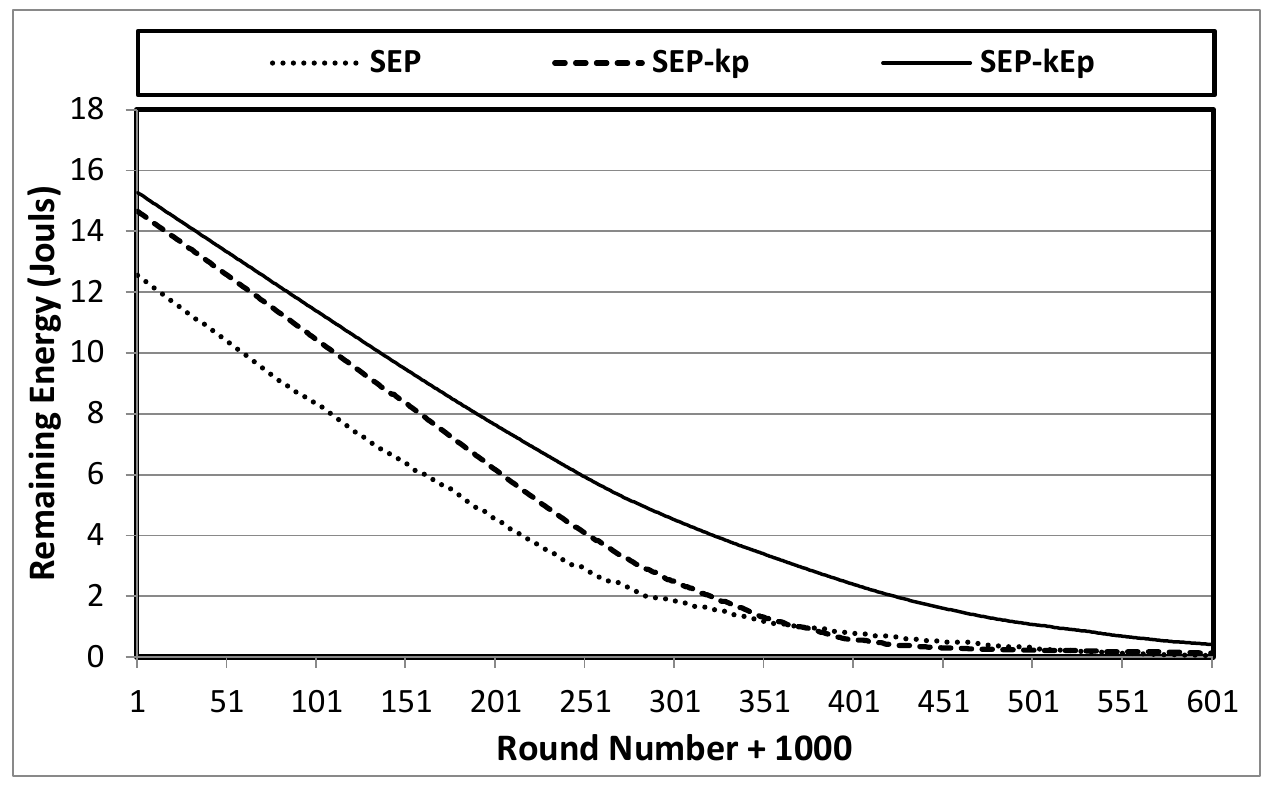}
	  \caption{\small{The Effect of $\kappa_{max}$, $E$, and $P_{adp}$ on Energy Consumption per Round in SEP-based Algorithms}}
      \label{fig:SEP_Comparison_remEnergyDead}
\end{figure}

The energy consumption performance of the proposed LEACH-based algorithms is compared in Figure~\ref{fig:LEACH_Comparison_remEnergyDead}, considering the effect of $\kappa_{max}$, $E$, and $P_{adp}$. Employing $P_{adp}$ decreases the total energy consumption because it increases the probability of qualified SNs to become cluster-heads by the time. LEACH-$\kappa E\!P$ shows an energy-efficient performance because its cluster-head selection probability accounts for the energy $E$ factor. The same thing is applied for SEP algorithms as shown in Figure~\ref{fig:SEP_Comparison_remEnergyDead}.
\begin{figure}[!h]
\centering
\captionsetup{justification=centering}
	  \includegraphics[width=0.65\textwidth]{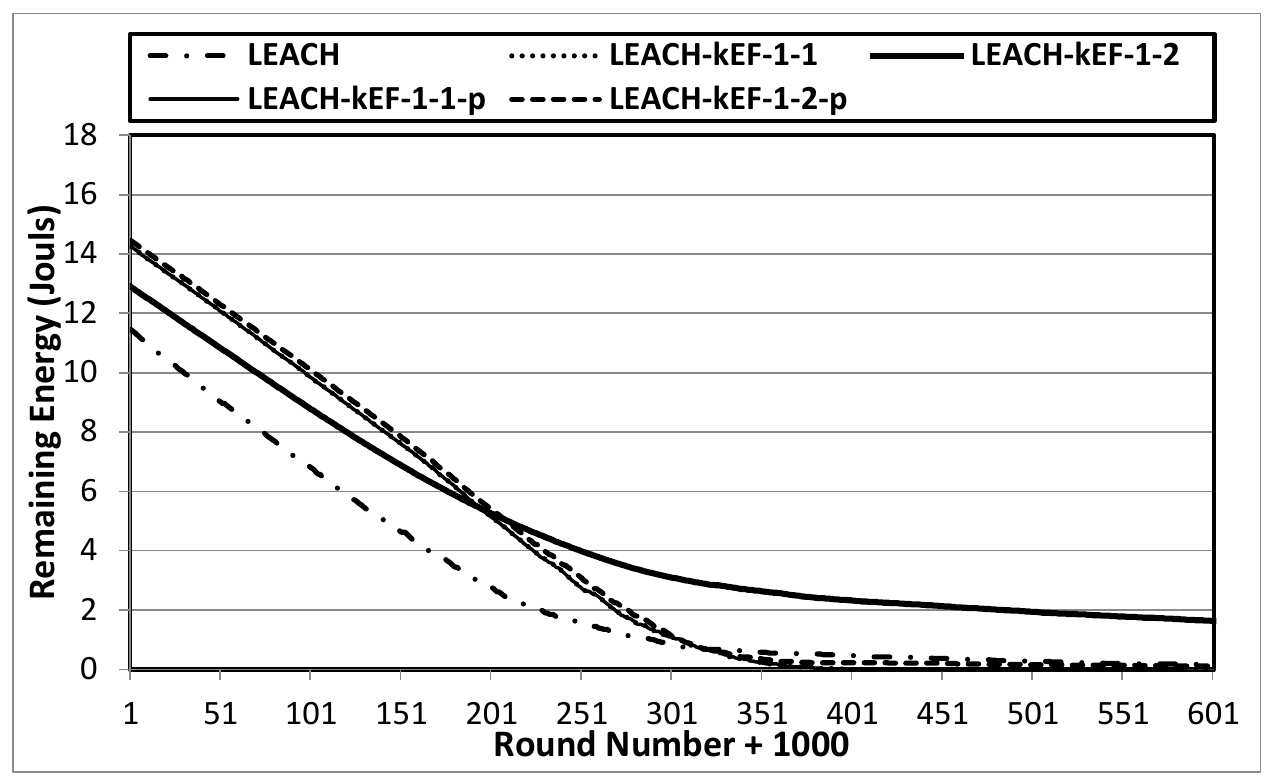}
	  \caption{\small{The Effect of $P_{adp}$ and $\digamma$ on Energy Consumption per Round in LEACH-based Algorithms}}
      \label{fig:LEACH_kEFpab_Comparison_remEnergyDead}
\end{figure}
\begin{figure}[!h]
\centering
\captionsetup{justification=centering}
	  \includegraphics[width=0.65\textwidth]{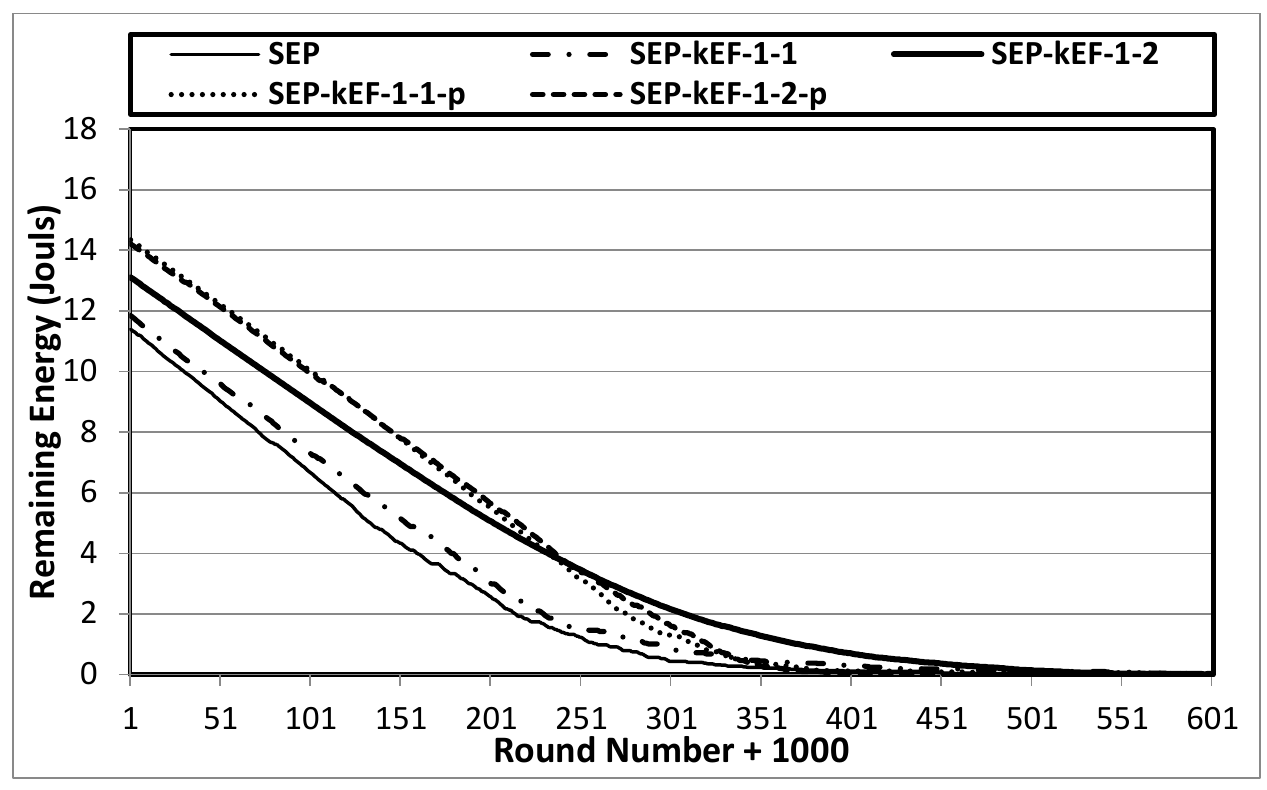}
	  \caption{\small{The Effect of $P_{adp}$ and $\digamma$ on Energy Consumption per Round in SEP-based Algorithms}}
      \label{fig:SEP_kEFpab_Comparison_remEnergyDead}
\end{figure}
\begin{figure}[!ht]
\centering
\captionsetup{justification=centering}
	  \includegraphics[width=0.65\textwidth]{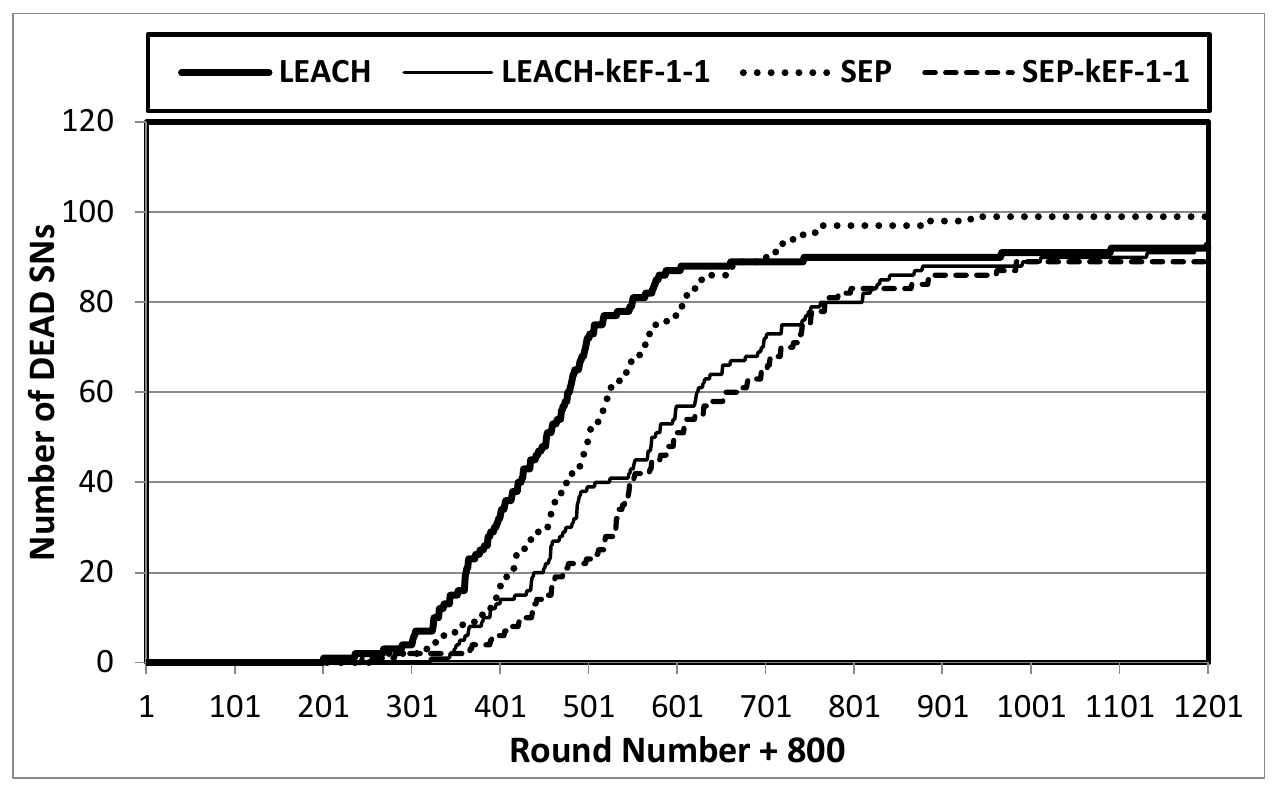}
	  \caption{\small{LEACH and SEP Algorithms without $P_{adp}$ for Number of Dead SNs per Round}}
      \label{fig:LEACH_SEP_withoutP}
\end{figure}
\begin{figure}[!ht]
\centering
\captionsetup{justification=centering}
	  \includegraphics[width=0.65\textwidth]{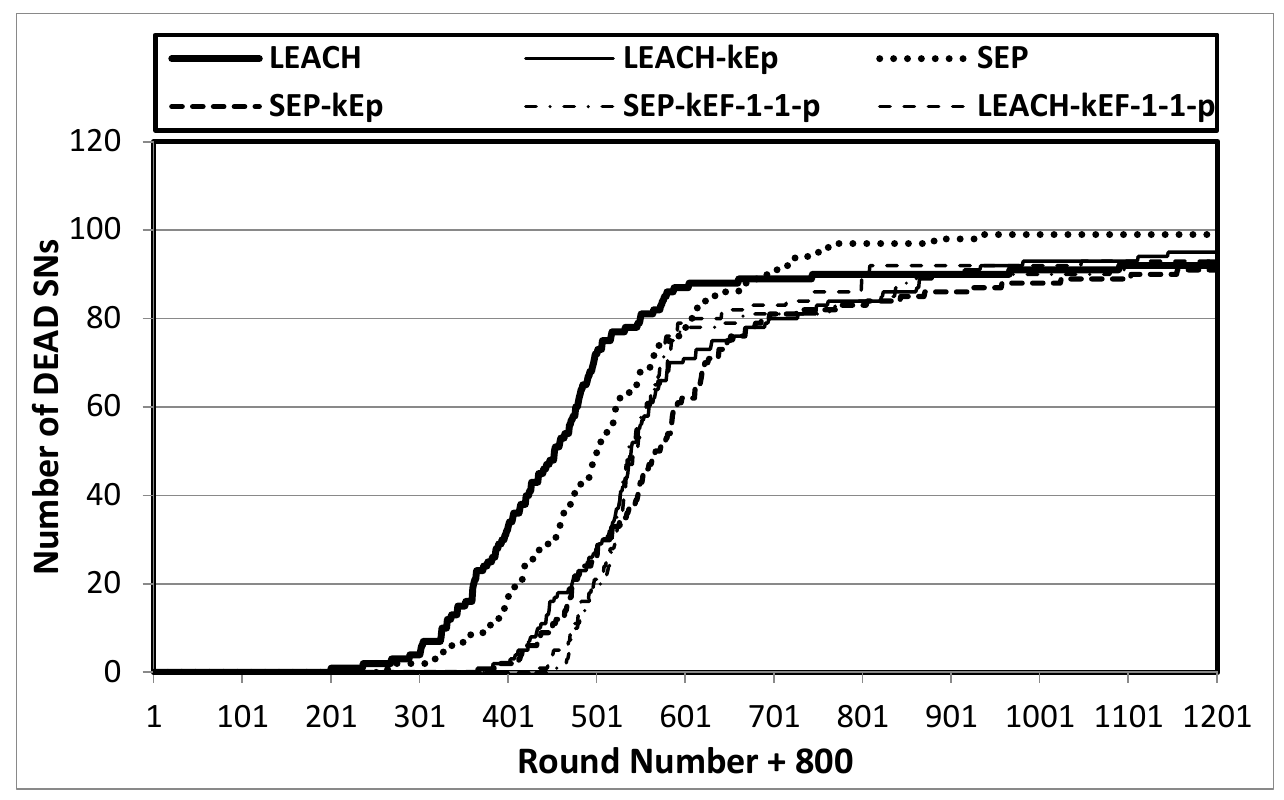}
	  \caption{\small{LEACH and SEP Algorithms with $P_{adp}$ for Number of Dead SNs per Round}}
      \label{fig:LEACH_SEP_withP}
\end{figure}

The impact of employing $\digamma$ in the algorithms is demonstrated in Figure~\ref{fig:LEACH_kEFpab_Comparison_remEnergyDead}. The adaptive LEACH-$\kappa E\!\digamma$\!-1-1-$P$ and LEACH-$\kappa E\!\digamma$\!-1-2-$P$ algorithms enhance the QoS performance higher than LEACH algorithm. The same thing is applied for SEP algorithms in Figure~\ref{fig:SEP_kEFpab_Comparison_remEnergyDead}, where adaptive SEP-$\kappa E\!\digamma$\!-1-1-$P$ and SEP-$\kappa E\!\digamma$\!-1-2-$P$ are the most energy-efficient algorithms.

To conclude, Figure~\ref{fig:LEACH_SEP_withoutP} assesses the performance of LEACH and SEP algorithms illustrating the best proposed algorithms according to probability of dead SNs, demonstrating the effect of $E$ and $\digamma$ performance measures. As well, Figure~\ref{fig:LEACH_SEP_withP} compares the mutual performance impact of such algorithms, utilizing $P_{adp}$ with $E$ and $\digamma$ measures. Overall, the proposed adaptive-based algorithms improve the probability of dead SNs and stability period of original LEACH and SEP algorithms.
\begin{table*}[!ht]
\label{tab:summary}
\centering
\begin{tabular}{cc|cc}
\hline
Algorithm                                   & Round Number & Algorithm                      & Round Number \\ \hline

LEACH                                         & 999   & SEP                                           & 1052 \\
LEACH-$\kappa\!P$                             & 1063  & SEP-$\kappa\!P$                               & 1081 \\
LEACH-$\kappa E\!P$                           & 1164  & SEP-$\kappa E\!P$                             & 1183 \\

LEACH-$\kappa E\!\digamma$\!-1-1              & 1135  & SEP-$\kappa E\!\digamma$\!-1-1                & 1150 \\
LEACH-$\kappa E\!\digamma$\!-1-2              & 1126  & SEP-$\kappa E\!\digamma$\!-1-2                & 1139 \\

LEACH-$\kappa E\!\digamma$\!-1-1-$P$          & 1230  & SEP-$\kappa E\!\digamma$\!-1-1-$P$            & 1241 \\
LEACH-$\kappa E\!\digamma$\!-1-2-$P$          & 1217  & SEP-$\kappa E\!\digamma$\!-1-2-$P$            & 1237 \\

LEACH-$\kappa E\!\digamma$\!-1-1-$P$-Learning & 1308  & SEP-$\kappa E\!\digamma$\!-1-1-$P$-Learning   & 1367 \\

LEACH-$\kappa E\!\digamma$\!-1-2-$P$-Learning & 1277  & SEP-$\kappa E\!\digamma$\!-1-2-$P$-Learning   & 1294  \\ \hline
\end{tabular}
\caption{The Round Number of the Death of First SN for each Clustering Algorithm}
\captionsetup{justification=centering}
\end{table*}

\subsection{Evolving $P_{adp}$ and $\kappa_{max}$}

\begin{figure}[!h]
\centering
\captionsetup{justification=centering}
	  \includegraphics[width=0.65\textwidth]{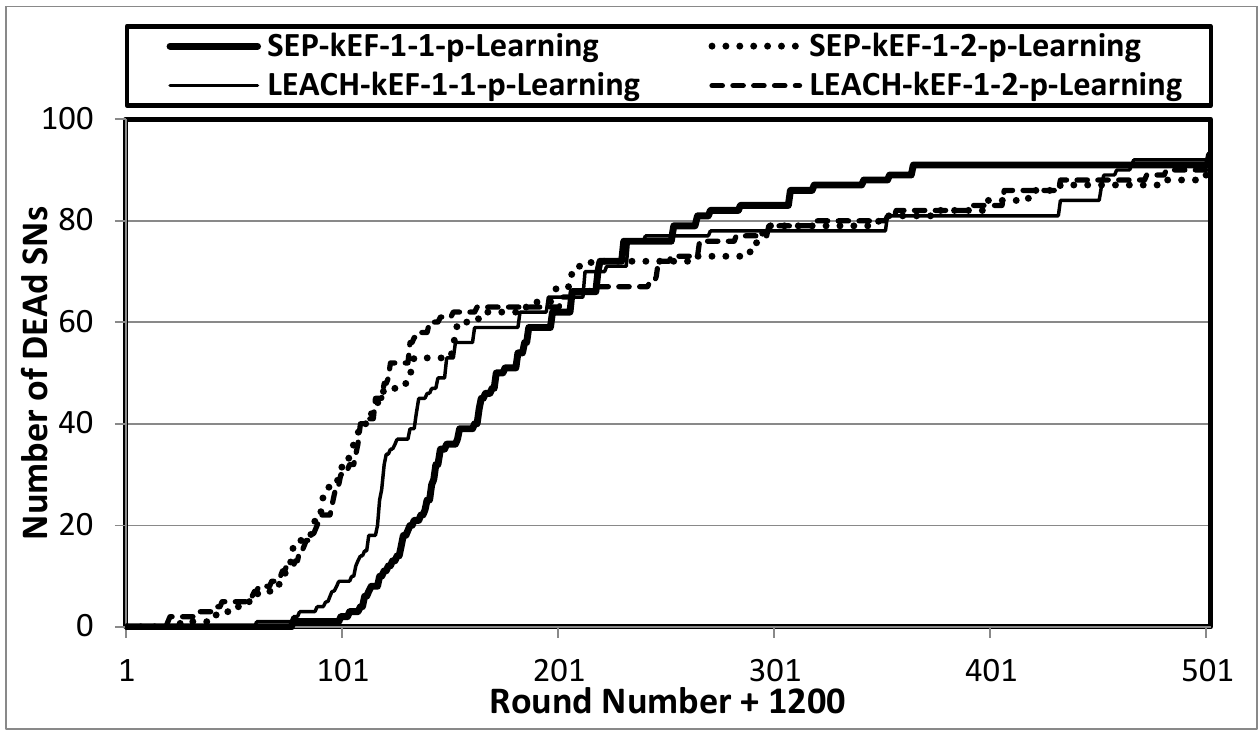}
	  \caption{\small{The Effect of $P_{adp}$ on Number of Dead SNs per Round in Learning-based LEACH and SEP Algorithms}}
      \label{fig:LEACH_SEP_Learning_top4}
\end{figure}

Learning $\kappa_{max}$-based algorithms are developed based on the former proposed algorithms, in which the clustering factor $\kappa_{max}$ is frequently evolved at the end of each round according to the probability of alive SNs $\zeta$, instead of having it fixed a priori in the proposed adaptive algorithms. Accordingly, the probability $P_{adp}$ is adapted according to the evolved $\kappa_{max}$ clustering and number of alive SNs $\zeta$ at the end of each round. Figure~\ref{fig:LEACH_SEP_Learning_top4} compares such learning-based algorithms. The round number period of comparison for the likelihood of dead SNs is chosen to be between $\nolinebreak{1200\!-\!1700}$, where the major difference between the learning algorithms appears.
\begin{figure}[!h]
\centering
\captionsetup{justification=centering}
	  \includegraphics[width=0.65\textwidth]{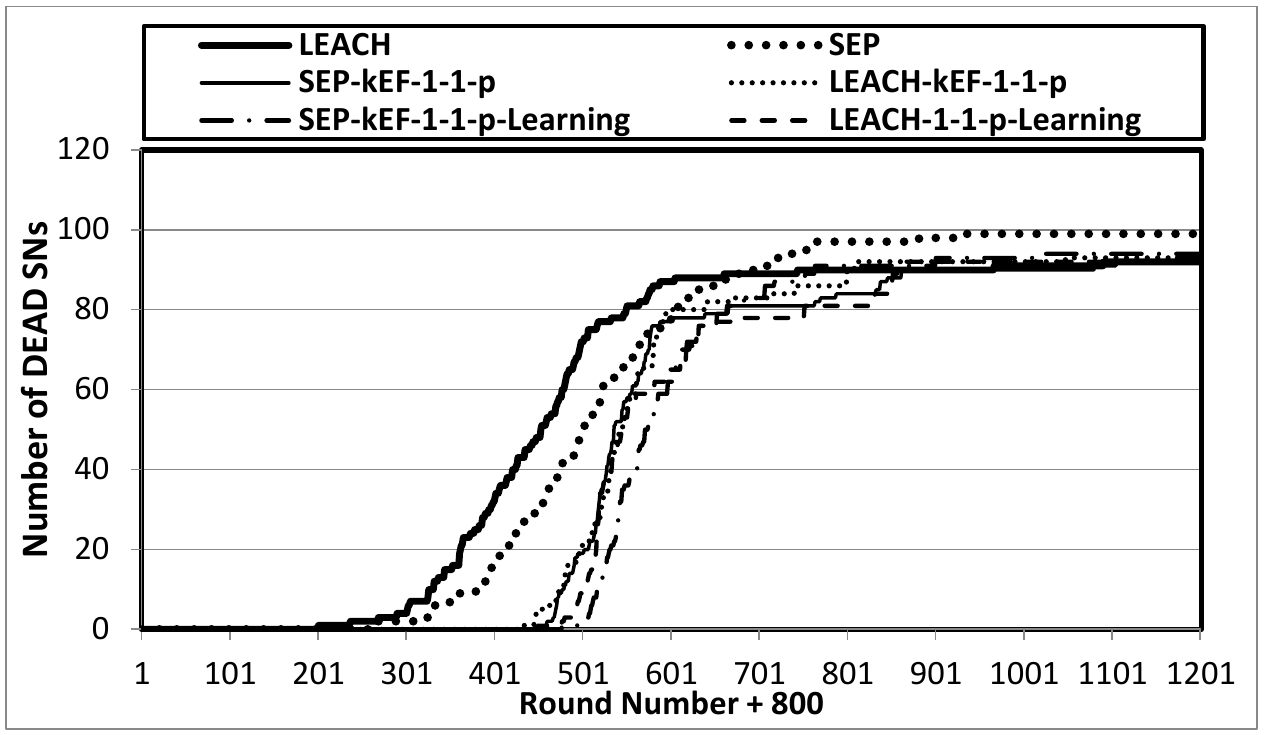}
	  \caption{\small{Learning-based LEACH and SEP Algorithms for Number of Dead SNs per Round, $\alpha\!=\!1$ and $\beta\!=\!1$}}
      \label{fig:LEACH_SEP_Learning_ALL11}
\end{figure}
\begin{figure}[!h]
\centering
\captionsetup{justification=centering}
	  \includegraphics[width=0.65\textwidth]{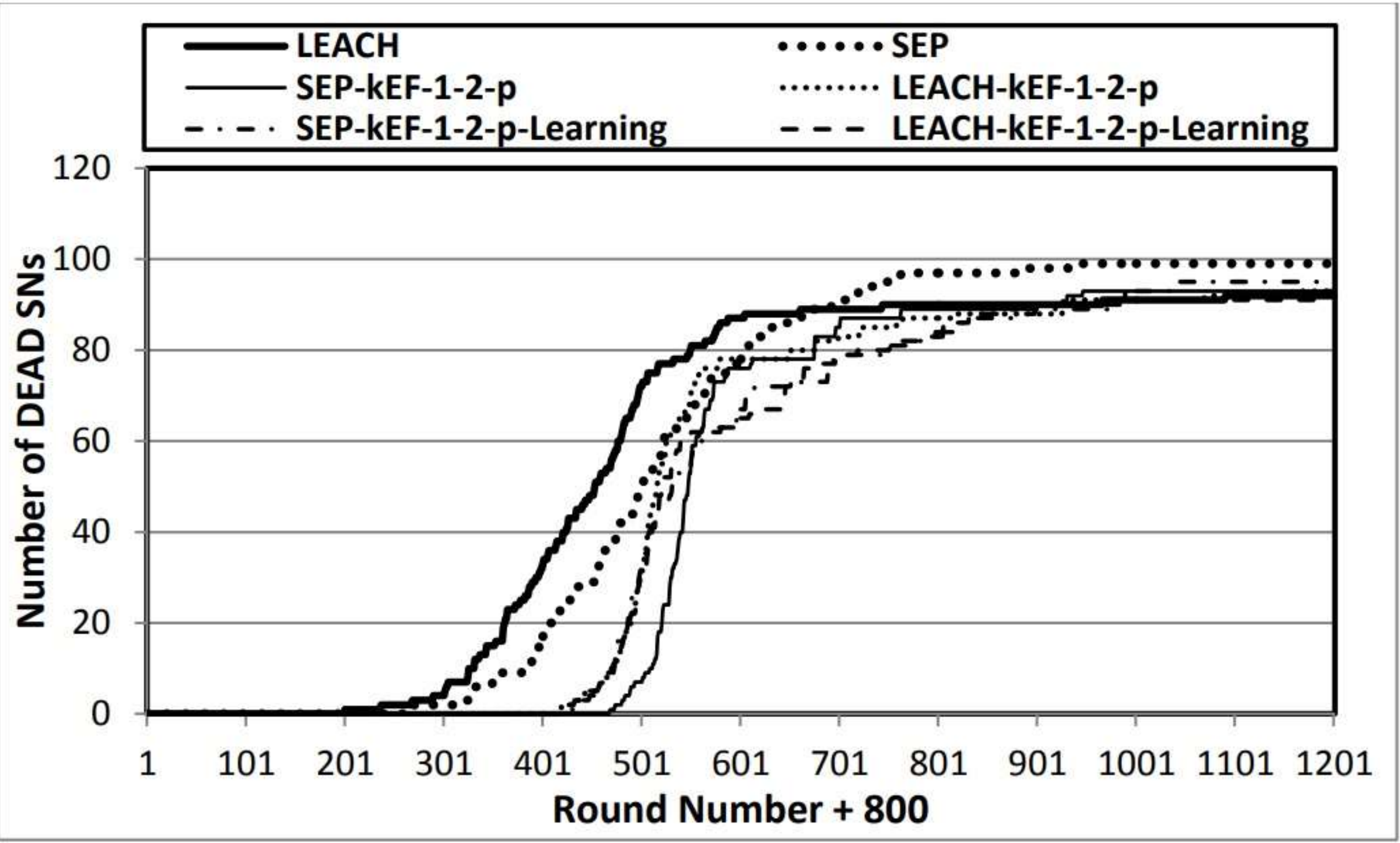}
	  \caption{\small{Learning-based LEACH and SEP Algorithms for Number of Dead SNs per Round, $\alpha\!=\!1$ and $\beta\!=\!2$}}
      \label{fig:LEACH_SEP_Learning_ALL12}
\end{figure}

The evaluation of the original LEACH and SEP algorithms, the proposed adaptive algorithms, and the learning $\kappa_{max}$-based adaptive algorithms are presented in Figures~\ref{fig:LEACH_SEP_Learning_ALL11}-\ref{fig:LEACH_SEP_Learning_top4}. The learning-based adaptive algorithms demonstrate better performance and improve the stability period, as compared to both the proposed adaptive and the original LEACH and SEP algorithms. In this case, the round number period of comparison for the likelihood of dead SNs is chosen to be between $\nolinebreak{800\!-\!2000}$, where the major difference between the proposed adaptive and learning algorithms appears. To summarize, Table~III presents the stability period represented by death of the first SN for each clustering algorithm. The overall QoS performance is improved when $\kappa_{max}$, $E$, $P_{adp}$, and $\digamma$ are employed together with learning.

\section{Conclusion}
\label{sec:conc}

It is observed that incorporating the residual energy $E$ of a SN in the cluster-head selection probability and limiting the number of cluster-heads permitted per round to $\kappa_{\!{max}}$ improve the overall QoS performance, for LEACH and SEP algorithms. This, however, decreases the likelihood of dead SNs and the total energy consumption, which as a result prolongs the lifetime of network operation. In addition, it is found that $P_{\!adp}$ improves the cluster-head selection probability because it is adaptive to existing network's state; represented by number of remaining SNs $\zeta$ per round and maximum number of clusters $\kappa_{\!{max}}$ permitted per round. Adapting the probability of selecting a cluster-head by $P_{\!adp}$ and limiting the number of permitted cluster-heads per round to $\kappa_{\!{max}}$ would both distribute the energy and defer qualified SNs evenly throughout the network operation, as well as improve the stability period and QoS performance. Such improvements involve probability of dead SNs and total energy consumption. In addition, incorporating the development of enhanced $\digamma$ ratio improves the performance by considering the residual energy of a cluster-head relative to its distances with its members, instead of only accounting for the distance to compose clusters. Such a relation accounts for the mutual performance impact of member displacements from their corresponding cluster-heads which in turn distributes existing SNs evenly among elected cluster-heads.

\section{Future Directions}
\label{sec:future}

The cluster-head selection probability will be modified to account for current SN's position to comprise a hierarchical distance-based clustering. Furthermore, a density-based clustering will be developed to formulate SN clusters that do not require a priori knowledge on $\kappa_{\!{max}}$. A sensitivity analysis will be conducted to study the effect and relationship between input and output performance parameters of the mathematical model. The study will mainly be focused on examining the effect of varying $\kappa_{\!{max}}$ on the probability of dead SNs per round. The QoS performance of the proposed algorithms will be compared with the original LEACH and SEP algorithms, by varying the level of heterogeneity and percentage of advanced SNs in the WSN.


%
\label{sect:bib}
\bibliographystyle{abbrv}
\bibliography{easychairWSN}

\section*{Author Biography}
\vspace*{1em}
\begin{biography}{Husam Suleiman}{husam} received his PhD in Electrical and Computer Engineering from University of Waterloo, Canada in 2019. His MSc degree is in Computer Engineering from Khalifa University, UAE in 2011 in collaboration with the Massachusetts Institute of Technology (MIT). His BSc degree is in Electrical and Computer Engineering from Hashemite University, Jordan in 2007. Currently, he is an assistant professor in Applied Science Private University, Jordan. His research interests include QoS Optimization in Multi-Tier Cloud Computing, Load Scheduling \& Balancing, Big Data Algorithms, Resource Allocation Methods, Performance Prediction Analysis, Security Requirements Engineering Methods for Smart Grids and Systems.
\end{biography}
\vspace*{0.5em}
\begin{biography}{Mohammad Hamdan}{mohamed} received his PhD degree in Electrical Engineering from University of Southampton, Southampton, UK in 2019. His MSc degree is in Energy, Sustainability with Electric Power Engineering from University of Southampton, UK in 2014. His BSc degree is in Electrical Engineering from Jordan University of Science and Technology, Irbid, Jordan in 2010. Currently, he is an assistant professor in Applied Science Private University, Jordan. His research interests include smart grid and renewable energy applications.
\end{biography}

%

\end{document}